\documentclass{jfm}  %[lineno]
\usepackage{graphicx}
\usepackage{overpic}
\usepackage{tikz}
\tikzset{>=stealth}
\graphicspath{{figures/}}
\usepackage{algorithm,algorithmic}
\usepackage{newtxtext}
\usepackage{newtxmath}
\usepackage{natbib}
\usepackage{hyperref}
\hypersetup{
    colorlinks = true,
    urlcolor   = blue,
    citecolor  = black,
}

\newcommand{\RomanNumeralCaps}[1]

\title{Deep reinforcement learning for tracking a moving target in jellyfish-like swimming}

\author{Yihao Chen\aff{1},
 \and Yue Yang\aff{2}\corresp{\email{yyg@pku.edu.cn}}}

\affiliation{\aff{1}State Key Laboratory for Turbulence and Complex Systems, College of Engineering, Peking University, Beijing 100871, China
\aff{2}HEDPS-CAPT, Peking University, Beijing 100871, China}

\begin{document}
\maketitle

\begin{abstract}
We develop a deep reinforcement learning method for training a jellyfish-like swimmer to effectively track a moving target in a two-dimensional flow. This swimmer is a flexible object equipped with a muscle model based on torsional springs. 
We employ a deep Q-network (DQN) that takes the swimmer's geometry and dynamic parameters as inputs, and outputs actions which are the forces applied to the swimmer. In particular, we introduce an action regulation to mitigate the interference from complex fluid-structure interactions. 
The goal of these actions is to navigate the swimmer to a target point in the shortest possible time. 
In the DQN training, the data on the swimmer's motions are obtained from simulations conducted using the immersed boundary method.  
During tracking a moving target, there is an inherent delay between the application of forces and the corresponding response of the swimmer's body due to hydrodynamic interactions between the shedding vortices and the swimmer's own locomotion. 
Our tests demonstrate that the swimmer, with the DQN agent and action regulation, is able to dynamically adjust its course based on its instantaneous state. 
This work extends the application scope of machine learning in controlling flexible objects within fluid environments. 
\end{abstract}

\section{Introduction}
\label{sec:intro}

Machine learning for fluid mechanics has garnered considerable attention in recent years \citep{annurev2020,Karnidakis2021physic}.
Reinforcement learning, one of the machine learning methods, employs an agent that learns to interact with its environment by taking actions and learning from the rewards it receives, with the aim of maximizing the cumulative reward~\citep{richard2018RL}. 
It has been successfully applied in fluid mechanics, such as swimming strategy \citep{zhao2020RL}, tracking \citep{Mirzakhanloo2020cloak}, schooling strategy \citep{koumou2016school}, and turbulence modeling \citep{koumou2022turbRL}.

The classic reinforcement learning uses the iteration method, dynamic programming, Monte--Carlo method and tabular method, whereas the deep reinforcement learning (DRL) uses a deep neural network~\citep{GARNIER2021DRLreview}.  
The DRL is suitable for high-dimensional or non-linear tasks with very large number of states, as it avoids modeling complex system and exploits the feature extraction capabilities of deep neural network.
It has been successfully applied to various control tasks such as walking \citep{Haarnoja2019robot,Su2023RLwakling,Itahashi2024RLwalking}, flying \citep{Reddy2016RLflying,beckerehmck2020learning,apolito2021RLflying} and swimming \citep{Rodwell2023torque,CHEN2022RLswim,verma2018RLswim} to enhance propulsion efficiency or reduce drag, and the active flow control for flows past cylinders  \citep{han2022cylinder,Fan2020cyliner,CHEN2024cylinder}. 

Specifically, for walking task, the DRL receives the joint angles and sensor readings of a minitaur robot and gives proper motor positions to make it walk efficiently~\citep{Haarnoja2019robot}. 
For flight control, the DRL controls a drone to marked position based on sensor data~\citep{beckerehmck2020learning}. 
For swimming task, the DRL ascertains the torque acting on the mass centre of a Joukowski aerofoil, enabling it to follow a predefined path~\citep{Rodwell2023torque}. 
For drag reduction, the DRL takes the surrounding flow velocity of a 2D cylinder as input and determines the optimal angular velocity that minimizes drag torque on the surface~\citep{han2022cylinder}. 

The DRL applications are based on various reinforcement learning methods, e.g.~the actor-critic algorithm \citep{Konda1999ActorCriticA} and deep Q-learning \citep{DQN2015}, and with various neural network structures, e.g.~the multilayer perceptron \citep{rosenblatt1962principles,HORNIK1989}, recurrent neural network \citep{rumelhart1986learning}, long short-term memory \citep{LSTM1997} and transformer \citep{transformer2017}. 

These studies have highlighted the potential of machine learning techniques, particularly DRL, in enhancing fluid dynamic performance through active flow control \citep{Fan2020cyliner,Taira2020control,han2022cylinder,Cylinder2022,FlowControl2023,Rodwell2023torque,Koumou2024drag, Kubo2022control,Vignon2023flowcontrol, Xie2023control}.
However, most of these applications primarily focused on relatively straightforward tasks, where the actions can swiftly exert their effects on the controlled object, with minor influences from fluid-structure interaction (FSI). 

Therefore, the DRL can be extended to the complex flow control problems with significant FSI effects.
Jellyfish is believed to have one of the highest locomotion efficiency among natural creatures \citep{Brad2013passive,Costello2021efficent}, which involves body deformation and strong interaction with surrounding flow. 
Previous experiments and simulations found that the jellyfish adopts an optimal locomotion of swimming forward and consuming less energy \citep{Nicole2020propulsion, KB2012Modeling, FJ2018Thrust, Matharu2022jellyZ}. 
Its swimming pattern and morphology are relatively simple, allowing for good maneuverability through controlling the swimming direction \citep{hoover2015jelly}.

Extensive research has been conducted on the forward propulsion mechanisms of jellyfish \citep{dular2009forward,hoover20173djelly,Miles2019DontBJ,miles2019jelly}, whereas investigations into their turning capabilities remain relatively scarce \citep{jelly2019}.  
The mechanism of forward propulsion is the formation and shedding of symmetric vortex rings with symmetric muscle contraction \citep{hoover20173djelly}, which is also similar to the flapping flight mechanism observed in natural birds \citep{WZN2010flap, WZN2012hover}.
The mechanism of turning involves two sets of nerve nets that control different parts of the muscle, enabling the asynchronous contraction of muscles \citep{jelly2019, hoover2021pnas}. 
For a model jellyfish swimming in a 2D flow, the mechanism of forward propulsion has been also well studied, which involves vortex shedding with acting force on certain part of the jellyfish muscle with parameter studies on the jellyfish's diameter, contraction frequency and tentacle \citep{hoover2015jelly, Miles2019DontBJ,miles2019jelly}. 

Here we study the turning of a jellyfish-like swimmer in 2D flows. Once both forward propulsion and turning are accomplished, it is possible for the jellyfish to navigate. 
In particular, we propose a challenging DRL task wherein we control the motion of a flexible swimmer that can both swim forward and turn to track specific targets. 
The locomotion of jellyfish typically involves strong FSI and deformation of object. Therefore the control of its motion can be challenging due to the intricate influence of wake vortices generated by its previous action.

The outline of this paper is as follows.
In \S\,\ref{sec:IBM_and_RL}, we introduce the methods of computational fluid dynamics (CFD) and reinforcement learning, and describe setups of simulation cases and tracking tasks. 
In \S\S\,\ref{sec:fix_target} and \ref{sec:moving_target}, we demonstrate the capability of the DRL on tracking fixed and moving target points in jellyfish-like swimming, respectively. 
Some discussions are presented in \S\,\ref{sec:conclusions}. 

\section{Deep reinforcement learning}
\label{sec:IBM_and_RL}

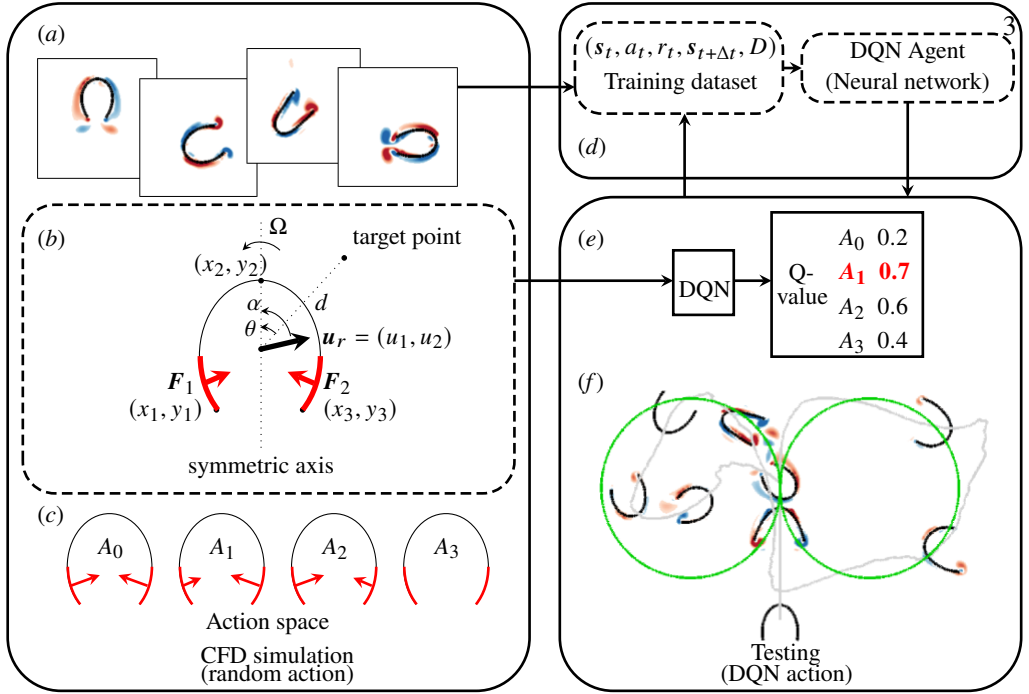
\begin{figure}
	\centering
	\begin{overpic}[scale=0.5]{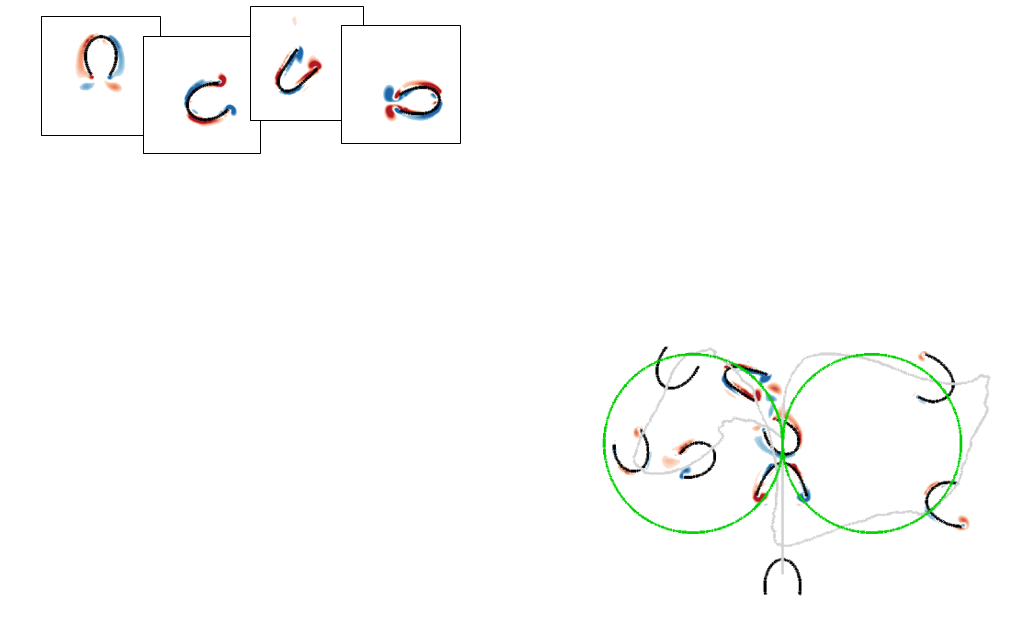} %big_pic7.png
		\put(82.5,39.7){%
			\begin{minipage}{0.4\textwidth} % 
				\color{red} % 
				\fontsize{9}{14}\selectfont %   
				$\boldsymbol{A_{1}}\,\,\,\boldsymbol{0.7}$ %  
			\end{minipage}
		}
		\put(82.5,43){%
			\begin{minipage}{0.4\textwidth} %
				\fontsize{9}{14}\selectfont
				$A_0\,\,\,0.2$
			\end{minipage}
		}
		\put(82.5,36.3){%
			\begin{minipage}{0.4\textwidth} %
				\fontsize{9}{14}\selectfont %
				$A_2\,\,\,0.6$
			\end{minipage}
		}
		\put(82.5,33){%
			\begin{minipage}{0.4\textwidth} %
				\fontsize{9}{14}\selectfont %
				$A_3\,\,\,0.4$
			\end{minipage}
		}
		\put(77.7,39.7){%
			\begin{minipage}{0.4\textwidth} %
				\fontsize{9}{14}\selectfont %
				Q-
			\end{minipage}
		}
		\put(76.5,37.5){%
			\begin{minipage}{0.4\textwidth} %
				\fontsize{9}{14}\selectfont %
				value
			\end{minipage}
		}
		\put(66.9,38){%
			\begin{minipage}{0.4\textwidth} %
				\fontsize{9}{14}\selectfont %
				DQN
			\end{minipage}
		}
		\put(71,0.5){%
			\begin{minipage}{0.4\textwidth} %
				\fontsize{9}{14}\selectfont %
				(DQN action)
			\end{minipage}
		}
		\put(74,2.5){%
			\begin{minipage}{0.4\textwidth} %
				\fontsize{9}{14}\selectfont %
				Testing
			\end{minipage}
		}
		\put(20,2.5){%
			\begin{minipage}{0.4\textwidth} %
				\fontsize{9}{14}\selectfont %
				CFD simulation
			\end{minipage}
		}
		\put(20,0.5){%
			\begin{minipage}{0.4\textwidth} %
				\fontsize{9}{14}\selectfont %
				(random action)
			\end{minipage}
		}
		\put(57,43){%
			\begin{minipage}{0.4\textwidth} %
				\fontsize{9}{14}\selectfont %
				(\textit{e})
			\end{minipage}
		}
		\put(20.5,5.5){%
			\begin{minipage}{0.4\textwidth} %
				\fontsize{9}{14}\selectfont %
				Action space
			\end{minipage}
		}
		\put(57,29){%
			\begin{minipage}{0.4\textwidth} %
				\fontsize{9}{14}\selectfont %
				(\textit{f})
			\end{minipage}
		}
		\put(57,52){%
			\begin{minipage}{0.4\textwidth} %
				\fontsize{9}{14}\selectfont %
				(\textit{d})
			\end{minipage}
		}
		\put(4,63){%
			\begin{minipage}{0.4\textwidth} %
				\fontsize{9}{14}\selectfont %
				(\textit{a})
			\end{minipage}
		}
		\put(4,43){%
			\begin{minipage}{0.4\textwidth} %
				\fontsize{9}{14}\selectfont %
				(\textit{b})
			\end{minipage}
		}
		\put(4,16){%
			\begin{minipage}{0.4\textwidth} %
				\fontsize{9}{14}\selectfont %
				(\textit{c})
			\end{minipage}
		}
		\put(14,8){%
			\begin{tikzpicture}[overlay,scale=0.7]
				\draw (0,0) arc (-45:225:0.8 and 1);
				\draw[line width=1pt, color=red] (0,0) arc (-45:0:0.8 and 1);
				\draw[line width=1pt, color=red] (-1.12,0) arc (225:180:0.8 and 1);
				\draw[line width=1pt, color=red][->](-1.3,0.35)--(-0.75,0.55);
				\draw[line width=1pt, color=red][->](0.15,0.35)--(-0.4,0.55);
				\node[above] at (-0.55,0.7){$A_0$};
			\end{tikzpicture}
		}
		\put(25,8){%
			\begin{tikzpicture}[overlay,scale=0.7]
				\draw (0,0) arc (-45:225:0.8 and 1);
				\draw[line width=1pt, color=red] (0,0) arc (-45:0:0.8 and 1);
				\draw[line width=1pt, color=red] (-1.12,0) arc (225:180:0.8 and 1);
				\draw[line width=1pt, color=red][->](-1.3,0.35)--(-0.95,0.5);
				\draw[line width=1pt, color=red][->](0.15,0.35)--(-0.4,0.55);
				\node[above] at (-0.55,0.7){$A_1$};
			\end{tikzpicture}
		}
		\put(36,8){%
			\begin{tikzpicture}[overlay,scale=0.7]
				\draw (0,0) arc (-45:225:0.8 and 1);
				\draw[line width=1pt, color=red] (0,0) arc (-45:0:0.8 and 1);
				\draw[line width=1pt, color=red] (-1.12,0) arc (225:180:0.8 and 1);
				\draw[line width=1pt, color=red][->](-1.3,0.35)--(-0.75,0.55);
				\draw[line width=1pt, color=red][->](0.15,0.35)--(-0.2,0.5);
				\node[above] at (-0.55,0.7){$A_2$};
			\end{tikzpicture}
		}
		\put(47,8){%
			\begin{tikzpicture}[overlay,scale=0.7]
				\draw (0,0) arc (-45:225:0.8 and 1);
				\draw[line width=1pt, color=red] (0,0) arc (-45:0:0.8 and 1);
				\draw[line width=1pt, color=red] (-1.12,0) arc (225:180:0.8 and 1);
				\node[above] at (-0.55,0.7){$A_3$};
			\end{tikzpicture}
		}
		
		\put(30,27){%
			\begin{tikzpicture}[overlay]
				\draw (0,0) arc (-45:225:0.8 and 1);
				\draw [dotted] (-0.55,-0.5)--(-0.55,2.5);
				\draw [dotted] (-0.55,0.8)--(0.55,2);
				\draw (-0.35,1)[->] arc (45:90:0.3);
				\node[above right] at (-0.88,0.85){$\theta$};
				\draw (-0.16,0.97)[->] arc (10:90:0.4);
				\node[above right] at (-0.85,1.2){$\alpha$};
				\draw (-0.3,2.1)[->] arc (45:135:0.35);
				\node[above right] at (-0.55,2.2){$\Omega$};
				\node[above right] at (0.55,2){target point};
				\node[below] at (-0.55,-0.5){symmetric axis};
				\node[right] at (0.15,0){$(x_{3},y_{3})$};
				\node[left] at (-1.2,0){$(x_{1},y_{1})$};
				\node[above] at (-0.95,1.65){$(x_{2},y_{2})$};
				\filldraw (-1.14,0) circle (0.03);
				\filldraw (0,0) circle (0.03);
				\filldraw (-0.55,1.7) circle (0.03);
				\filldraw (0.55,2) circle (0.03);
				\filldraw (-0.55,0.8) circle (0.03);
				\node[above] at (0.25,1.2){$d$};
				\draw[line width=2pt, color=red] (0,0) arc (-45:0:0.8 and 1);
				\draw[line width=2pt, color=red] (-1.12,0) arc (225:180:0.8 and 1);
				\draw[line width=2pt, color=red][->](-1.3,0.35)--(-0.95,0.5);
				\draw[line width=2pt, color=red][->](0.15,0.35)--(-0.2,0.5);
				\node[left] at (-1.3,0.35){$\boldsymbol{ F}_{1}$};
				\node[right] at (0.15,0.35){$\boldsymbol{ F}_{2}$};
				\draw[line width=2pt][->](-0.55,0.8)--(0.1,0.95);
				\node[right] at (0.15,0.95){$\boldsymbol{u}_r=(u_{1},u_{2})$};
				
				\draw[dash pattern=on 4pt off 2pt, rounded corners=12pt,line width=1pt] (-3.7,-1.1) rectangle (2.8,2.7);
				\draw[line width=1pt][->](2.8,1.7)--(4.9,1.7);
				\draw[line width=1pt] (4.9,1.25) rectangle (5.7,2.1);
				\draw[line width=1pt][->](5.7,1.7)--(6.2,1.7);
				\draw[line width=1pt] (6.2,0.7) rectangle (8.2,2.6);
				\draw[rounded corners=20pt,line width=1pt] (3.4,-3.7) rectangle (9.5,2.8);
				\draw[line width=1pt][->](5.05,2.8)--(5.05,3.9);
				\draw[line width=1pt][->](8,4.05)--(8,2.8);
				\draw[dash pattern=on 4pt off 2pt, rounded corners=10pt,line width=1pt] (3.6,3.9) rectangle (6.35,5.1);
				\draw[dash pattern=on 4pt off 2pt, rounded corners=8pt,line width=1pt] (6.6,4.05) rectangle (9.4,4.95);
				\draw[line width=1pt][->](6.35,4.5)--(6.6,4.5);
				
				\node [above] at (5.0,4.5) {$(\boldsymbol{s}_{t},a_{t},r_{t},\boldsymbol{s}_{t+\Delta t},D)$};
				\node [below] at (5.0,4.55) {Training dataset};
				\node [above] at (8,4.45) {DQN Agent};
				\node [below] at (8,4.55) {(Neural network)};
				\draw[rounded corners=12pt,line width=1pt] (3.4,3.05) rectangle (9.5,5.35);
				\draw[line width=1pt][->](2.05,4.25)--(3.6,4.25);
				
				\draw[rounded corners=20pt,line width=1pt] (-3.9,-3.7) rectangle (3,5.35);
			\end{tikzpicture}%
		}
	\end{overpic}
	\caption{Diagram for the overall workflow. 
         (\textit{a}) Data obtained from multiple simulations are used for offline training. 
         (\textit{b}) Geometry and state of the jellyfish-like swimmer. The red parts indicate where the forces are applied. 
         (\textit{c}) Action space with four actions $A_i$, $i=0,1,2,3$, representing typical jellyfish actions (from left to right): symmetric forces on the two sides, larger force on the right side, larger force on the left side, and no force. 
         (\textit{d}) Multiple simulations with random actions (left dashed box) and collection of the experience tuple $(\boldsymbol{s}_{t},a_{t},r_{t},\boldsymbol{s}_{t+\Delta t},D)$ (right dashed box). 
         (\textit{e}) The DQN module receives the state vector and outputs a Q-value for each action. The action with the highest Q-value is finally chosen. 
         (\textit{f}) The trained agent is tested in various tracking tasks. Supplementary movies 1 and 2 illustrate how the swimmer tracks a moving target.}\label{fig:state_and_geo}
\end{figure}

The overall workflow is sketched in figure \ref{fig:state_and_geo}. We first run multiple simulations to collect data of jellyfish-like swimming (figure \ref{fig:state_and_geo}\textit{a}). The dataset is then used to train the neural network offline (figure \ref{fig:state_and_geo}\textit{d}). The network is tested in various tracking tasks (figure~\ref{fig:state_and_geo}\textit{f}). Supplementary movies 1 and 2 illustrate how the swimmer tracks a moving target. 
Moreover, figures \ref{fig:state_and_geo}\textit{b,c,e} show the jellyfish's state, action space and decision making process.

\subsection{Data preparation}
The 2D flow data for the deep reinforcement learning of jellyfish-like swimming are obtained from numerical simulations. The immersed boundary method \citep{peskin2002immersed,Tong2021IB} is used to treat the fluid-solid coupling at the moving boundary.
A unit density, incompressible flow is governed by the Navier--Stokes equations
\begin{eqnarray}
	\frac { \partial \boldsymbol { u } } { \partial t } + \boldsymbol { u } \cdot \nabla \boldsymbol { u } &=& - \nabla p + \nu \nabla ^ { 2 } \boldsymbol { u } + \boldsymbol { f },\label{eq:ns_eqn} \\
	\nabla \cdot  \boldsymbol { u }&=&0,\label{eq:consective_eqn}
\end{eqnarray}
where $\boldsymbol { u }$, $p$, $\nu$ and $\boldsymbol {f}$ denote the velocity, pressure, kinematic viscosity and body force exerted by a jellyfish-like swimmer.

The immersed boundary is represented by Lagrangian markers.
A regularised delta function $\delta_h$ \citep{peskin2002immersed} is employed to interpolate and spread $\boldsymbol {f}$ between Eulerian and Lagrangian points, whose coordinates are denoted by $\boldsymbol {x}$ and $\boldsymbol {X}$, respectively.
The Eulerian force in \eqref{eq:ns_eqn} is calculated as
\begin{equation}\label{eq: eulerian_lagrangian_force}
	\boldsymbol { f } ( \boldsymbol { x }) = \int_{S}  \boldsymbol { F } ( \boldsymbol { X }) \delta_h ( \boldsymbol { x } - \boldsymbol { X } ) d \boldsymbol{X},
\end{equation}
where $\boldsymbol{F}$ denotes the Lagrangian force at $\boldsymbol{X}$, and $S$ is the domain of the immersed boundary.
The non-slip condition is satisfied by exerting $\boldsymbol{F}$ on the immersed boundary.
The velocity on the immersed boundary satisfies
\begin{equation}\label{eq:no_slip_vel}
	\int _ { \mathcal{D} } \boldsymbol { u } ( \boldsymbol { x }  ) \delta_h (\boldsymbol { x } - \boldsymbol { X }) d \boldsymbol { x } = \boldsymbol { U } _ { b } ( \boldsymbol { X }),
\end{equation}
where $\mathcal{D}$ denotes the entire fluid domain and $\boldsymbol { U } _ { b }$ the velocity at Lagrangian points.

The simulations were conducted using the code IBAMR \citep{GRIFFITH2007IB}, which is a distributed-memory parallel implementation of the immersed boundary method with adaptive mesh refinement for the Cartesian grid.

\subsection{Deep Q-learning}
A reinforcement learning system consists of two parts -- the agent and the environment.
The agent observes the status of environment and takes an optimal action based on this observation.
The environment then changes and the agent receives a reward by evaluating the action.
These steps loop until the end of a learning process. 

The deep Q-learning \citep{DQN2015}, a reinforcement learning method, uses the deep Q-network (DQN) to generate the optimal action. 
This neural network receives state $s_{t}$ at time $t$ as input, and outputs a value $Q(s_{t},A_i)$ for each action $A_i$, representing the highest total reward for the agent after taking action $A_i$.
This Q-value satisfies the Bellman optimality equation \citep{richard2018RL}
\begin{equation}\label{eq:bell_eqn}
	Q(s_{t},A_{i})=r_{t}+\mathop{\mathrm{max}}_{A_{j}}  Q(s_{t+\Delta t},A_{j}).
\end{equation}
Namely, given state $s_{t}$ and action $A_{i}$, the maximum total reward at $t$ equals reward $r_{t}$ after taking this action plus the maximum total reward at the next state $s_{t+\Delta t}$ among all possible actions $A_{j}$, where $\Delta t$ denotes the time interval.
For clarity, we define $t^*= t/\Delta t$, along with current state $s_{t^*}$ and next state $s_{t^*+1}$.
Note that the number of available actions is finite in the DQN.
There are several difficulties in a reinforcement learning task -- proper forms of state vector and reward function, overall convergence for a Bellman optimality equation, and gap between the convergence and actual performance of the agent.

\subsection{Case setup and action regulation}
In our 2D flow simulation, we adopt the bell-like geometry shown in figure \ref{fig:state_and_geo}\textit{b} for jellyfish-like swimming.
The outflow boundary condition is applied to all boundaries of $\Omega$.
The domain size is sufficiently large so that the boundary conditions do not influence the simulation results.

In the jellyfish-like locomotion, a pair of sinusoidal forces is applied to the swimmer's tips (marked in red), as illustrated in figure \ref{fig:state_and_geo}\textit{b}.
%
%The force directions are symmetric about the axis of symmetry of the bell (dash-dotted line).
%
The force density (force per unit length)
\begin{equation}
	\boldsymbol{F}_{i}= F_i\,\mathrm{sin}(2\pi ft)\boldsymbol{\tau}_{i},~~~i=1,2,
\end{equation}
acting on the left and right halves of the swimmer are denoted as $\boldsymbol{F}_1$ and $\boldsymbol{F}_2$, respectively, with $F_i = |\boldsymbol{F}_{i}|$ and the unit tangent vector $\boldsymbol{\tau}_{i}$ on $S$.
As sketched in figure~\ref{fig:state_and_geo}\textit{c}, $F_1$ and $F_2$ can be different, while they have the same frequency $f$ and zero initial phase.
Other setup details are similar to those in \citet{hoover2015jelly} except that the model for the force density due to the elastic deformation of the swimmer is improved, which is detailed in Appendix \ref{sec:elasitc_force}. The simulation parameters are listed in table \ref{table:para}. All parameters are non-dimensionalized by the reference length $l=1$ m and reference time 1 s. The diameter $d_0$ of the jellyfish-like swimmer is defined as the distance between the left and right ends of the swimmer at rest.

\begin{table}
	\begin{center}
		\setlength{\tabcolsep}{35pt}
		\begin{tabular}{ccc}
			%Parameter & Symbol & Value   \\
			Swimmer diameter  & $d_{0}$ & 0.1 \\
			Domain length & $W$  & 1.6\\
			Frequency & $f$  & 0.5\\
			Period & $T=1/f$ & 2\\
			Kinematic viscosity & $\nu$ & $5\times10^{-5}$ \\
			Characteristic velocity & $U=fd_0$ & 0.05\\
			Reynolds number & $\mathrm{Re} = Ul/\nu$ & 100\\
		\end{tabular}
		\caption{Parameters in the simulation for jellyfish-like swimming.}
		\label{table:para}
	\end{center}
\end{table}

As listed in table \ref{table:regulation}, we divide a period $t = 0 \sim T$ of the sinusoidal force pair $(F_1,F_2)$ applied to the swimmer into four quarters, with $T=1/f$. 
The force pair is non-dimensionalized by the reference force density 1 $\mathrm{N}/\mathrm{m}$.
During each quarter, a DQN is employed to process the current state and produce $(F_1,F_2)$.
In order to reduce the training complexity, we restrict the choices for $(F_1,F_2)$ to $(0.003,0.003)$, $(0.001,0.003)$, $(0.003,0.001)$, and $(0,0)$. These choices are labeled as actions $A_i$ with $i=0$, 1, 2, and 3, as listed in table \ref{table:regulation} and illustrated in figure \ref{fig:state_and_geo}\textit{c}. 
To mimic the jellyfish locomotion and achieve high propulsion efficiency and maneuverability, forces are applied only during the first and third quarters of each period, except for $A_0$ and $A_3$ (with symmetric force magnitudes) which can last for the next quarter following specific actions.
The actions are summarised in table \ref{table:regulation}.
This action regulation is similar to the burst-and-coast strategy used in optimal intermittent swimming of fish \citep{li2021burstcoast} and jellyfish-like swimming \citep{kang2023burst}.

\begin{table}
	\begin{center}
		\setlength{\tabcolsep}{8pt}
		\begin{tabular}{cccccc}
			Action & $(F_1,F_2)$ & $0\sim T/4$ & $T/4\sim T/2$ & $T/2\sim 3T/4$ & $3T/4\sim T$  \\
			$A_0$  & $(0.003,0.003)$ & $\checkmark$ & $\checkmark$ (follow $A_0$) & $\checkmark$ & $\checkmark$ (follow $A_0$)\\
			$A_1$ & $(0.001,0.003)$ & $\checkmark$ & $\times$ & $\checkmark$ & $\times$ \\
			$A_2$ & $(0.003,0.001)$ & $\checkmark$ & $\times$ & $\checkmark$ & $\times$ \\
			$A_3$ & $(0,0)$ & $\checkmark$ & $\checkmark$ (follow $A_1$$\sim$$A_3$) & $\checkmark$ & $\checkmark$ (follow $A_1$$\sim$$A_3$) \\
		\end{tabular}
	\end{center}
	\caption{Action regulation: forces are applied only during the first and third quarters of each period, except for $A_0$ and $A_3$, which can last for the next quarter following specific actions.}
	\label{table:regulation}
\end{table}

To evaluate the efficiency of model performance, we use the total time taken for the swimmer to complete a task.
The task is considered ended when the swimmer reaches the target or the swimmer gets too close to the boundary of the simulation domain.

\subsection{Training}
As sketched in figure~\ref{fig:state_and_geo}\textit{b}, the swimmer's state is characterized by a 12-dimensional vector
\begin{equation}\label{eq:state_vec}
	\boldsymbol{s}=(x_{1},y_{1},x_{2},y_{2},x_{3},y_{3},u_{1},u_{2},d,\theta,\Omega,n),
\end{equation}
where $(x_{i},y_{i}),~i=1,2,3$ denote the coordinates of the swimmer's left, middle, and right endpoints referenced to its mass centre, $\boldsymbol{u}_r = (u_{1},u_{2})$ the velocity of the mass centre relative to target, $d$ the distance between the mass centre and target, $\theta$ the angle between the swimmer's symmetric axis and the line segment connecting the mass centre and target, $\Omega$ the angular velocity of the symmetric axis, and $n=0,1,2,3$ indicates the quarter of period.
Note that the symmetric axis is defined as the line connecting the midpoint of the swimmer and its mass centre, and it does not necessarily imply that the swimmer geometry is symmetric all the time.
The state vector in \eqref{eq:state_vec} encapsulates the information perceived by a swimming jellyfish when it chooses to navigate to a specific location within its natural environment, and the choice of state variables is important for the further training and tracking performance. 

We choose a function $r(\boldsymbol{s},a)=-\mathrm{min}(\theta^{2},3)+\mathcal A\,\mathrm{clip}(v,-0.1,0.1)-\mathcal B d$ to characterize the reward received at state $\boldsymbol{s}$ after taking action $a$, where $v$ is the projection of mass centre's velocity onto the line segment connecting the mass centre and target point, and constants $\mathcal A$ and $\mathcal B$ are tuned during the training; the clip function $\mathrm{clip}(x,b,c)$ returns $x$ if $b\le x\le c$, $b$ if $x<b$, and $c$ if $x>c$.
The DQN agent receives state and estimates $Q^{*}(\boldsymbol{s},a)$.
The loss function is defined by
\begin{equation}\label{eq:loss}
	\mathrm{loss} = (Q^{*}(\boldsymbol{s},a)-(r(\boldsymbol{s},a)+(1-D)\mathop{\mathrm{max}}_{A_{j}}  Q^{*}(\boldsymbol{ s}^{\prime},A_{j})))^{2},
\end{equation}
where $D=1$ if the task is ended and $0$ otherwise, $\boldsymbol{s}^\prime $ denotes the next state and its subscripts for time is omitted for clarity. This loss function optimizes the DQN parameters to satisfy the Bellman optimality equation in \eqref{eq:bell_eqn} with a slight modification. The term $(1-D)$ means that the Q-value at the penultimate state is equivalent to the reward obtained.
More details on reinforcement learning and training are provided in Appendix \ref{sec:training_details}.

\section{Tracking fixed target}\label{sec:fix_target}
\subsection{Tracking performance}
\begin{figure}
	\centering
	%\includegraphics[width=.99\linewidth]{fix_target_alter5.png}
	%\begin{overpic}[scale=0.5]{fix_target_alter6.png}
    \begin{overpic}[scale=0.5]{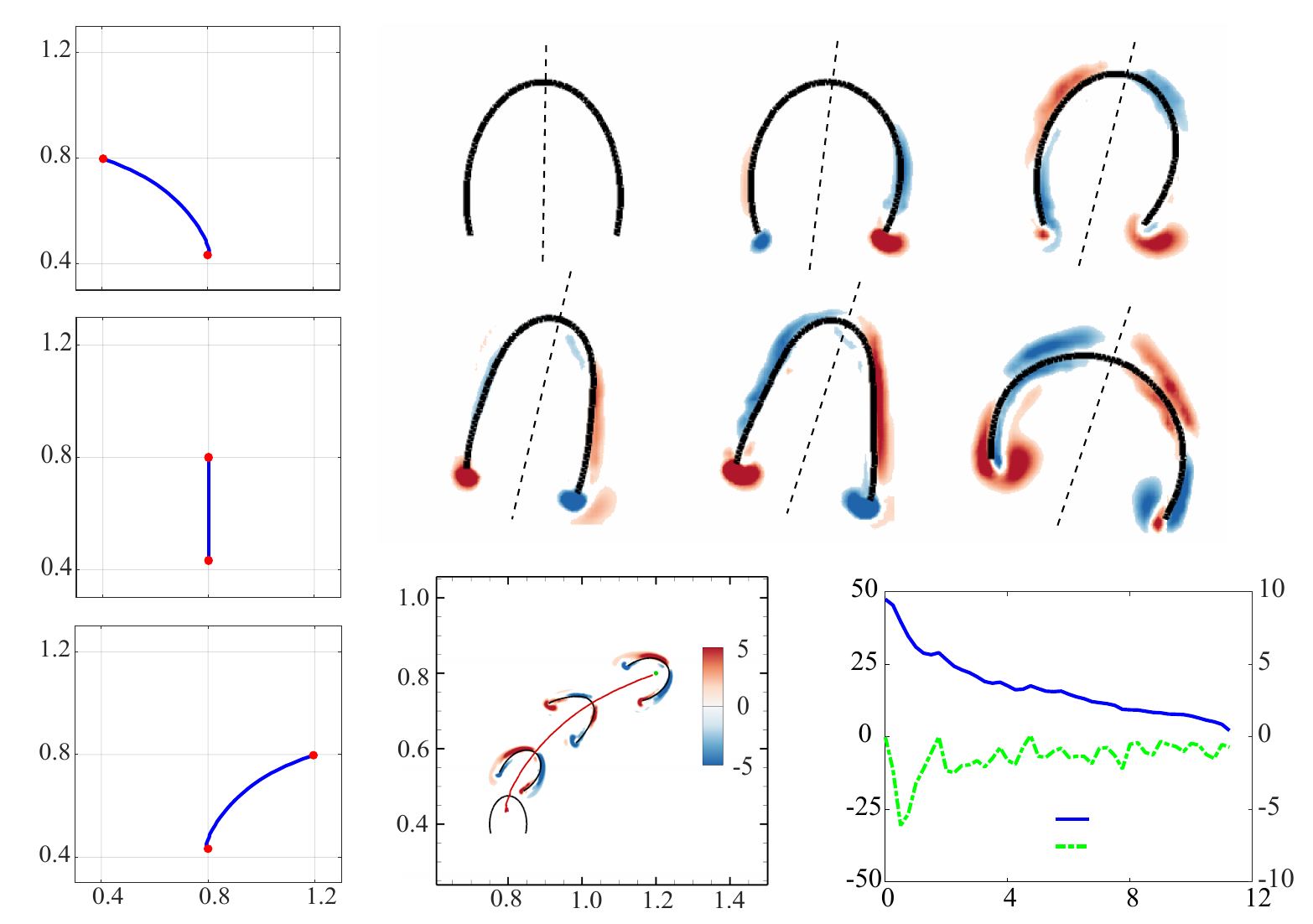}
		\put(91,46){%
			\begin{minipage}{0.4\textwidth}
				\fontsize{9}{14}\selectfont %   
				$t/T=1$ %  
			\end{minipage}
		}
		\put(91,43){%
			\begin{minipage}{0.4\textwidth}
				\fontsize{9}{14}\selectfont %   
				$a=A_0$ %  
			\end{minipage}
		}
		\put(91,62){%
			\begin{minipage}{0.4\textwidth}
				\fontsize{9}{14}\selectfont %   
				$t/T=0.4$ %  
			\end{minipage}
		}
		\put(91,59){%
			\begin{minipage}{0.4\textwidth}
				\fontsize{9}{14}\selectfont %   
				$a=A_3$ %  
			\end{minipage}
		}
		\put(69,62){%
			\begin{minipage}{0.4\textwidth}
				\fontsize{9}{14}\selectfont %   
				$t/T=0.2$ %  
			\end{minipage}
		}
		\put(69,59){%
			\begin{minipage}{0.4\textwidth}
				\fontsize{9}{14}\selectfont %   
				$a=A_1$ %  
			\end{minipage}
		}
		\put(47,62){%
			\begin{minipage}{0.4\textwidth}
				\fontsize{9}{14}\selectfont %   
				$t/T=0$ %  
			\end{minipage}
		}
		\put(47,59){%
			\begin{minipage}{0.4\textwidth}
				\fontsize{9}{14}\selectfont %   
				$a=A_1$ %  
			\end{minipage}
		}
		\put(47,46){%
			\begin{minipage}{0.4\textwidth}
				\fontsize{9}{14}\selectfont %   
				$t/T=0.7$ %  
			\end{minipage}
		}
		\put(47,43){%
			\begin{minipage}{0.4\textwidth}
				\fontsize{9}{14}\selectfont %   
				$a=A_0$ %  
			\end{minipage}
		}
		\put(69,46){%
			\begin{minipage}{0.4\textwidth}
				\fontsize{9}{14}\selectfont %   
				$t/T=0.8$ %  
			\end{minipage}
		}
		\put(69,43){%
			\begin{minipage}{0.4\textwidth}
				\fontsize{9}{14}\selectfont %   
				$a=A_0$ %  
			\end{minipage}
		}
		\put(47,46){%
			\begin{minipage}{0.4\textwidth}
				\fontsize{9}{14}\selectfont %   
				$t/T=0.7$ %  
			\end{minipage}
		}
		\put(66,48){%
			\begin{tikzpicture}[overlay,scale=0.7]
				\draw[line width=1.5pt, color=red][->](-5.75,1)--(-5,1.05);
				\draw[line width=1.5pt, color=red][->](-3.55,1)--(-4.55,1.1);
				\draw[line width=1.5pt, color=red][->](-1.65,1)--(-1,1.1);
				\draw[line width=1.5pt, color=red][->](0.4,1)--(-0.55,1.2);
				\draw[line width=1.5pt, color=red][->](-5.75,-2.4)--(-6.5,-2.25);
				\draw[line width=1.5pt, color=red][->](-4.1,-2.75)--(-3.35,-2.95);
				\draw[line width=1.5pt, color=red][->](-1.8,-2.35)--(-2.55,-2.05);
				\draw[line width=1.5pt, color=red][->](0.05,-2.75)--(0.75,-2.8);
				\draw[line width=1.5pt, color=red][->](1.8,-2.3)--(0.9,-2.15);
				\draw[line width=1.5pt, color=red][->](4.4,-3.15)--(5.25,-3.4);
			\end{tikzpicture}
		}
		\put(-1.5,67){%
			\begin{minipage}{0.4\textwidth}
				\fontsize{9}{14}\selectfont %   
				(\textit{a}) %  
			\end{minipage}
		}
		\put(-1.5,45){%
			\begin{minipage}{0.4\textwidth}
				\fontsize{9}{14}\selectfont %   
				(\textit{b}) %  
			\end{minipage}
		}
		\put(-1.5,23){%
			\begin{minipage}{0.4\textwidth}
				\fontsize{9}{14}\selectfont %   
				(\textit{c}) %  
			\end{minipage}
		}
		\put(32,67){%
			\begin{minipage}{0.4\textwidth}
				\fontsize{9}{14}\selectfont %   
				(\textit{d}) %  
			\end{minipage}
		}
		\put(32,28){%
			\begin{minipage}{0.4\textwidth}
				\fontsize{9}{14}\selectfont %   
				(\textit{e}) %  
			\end{minipage}
		}
		\put(63,28){%
			\begin{minipage}{0.4\textwidth}
				\fontsize{9}{14}\selectfont %   
				(\textit{f}) %  
			\end{minipage}
		}
		\put(1,57){%
			\begin{minipage}{0.4\textwidth}
				\fontsize{9}{14}\selectfont %   
				$\boldsymbol{\textit{y}}$ %  
			\end{minipage}
		}
		\put(1,34.5){%
			\begin{minipage}{0.4\textwidth}
				\fontsize{9}{14}\selectfont %   
				$\boldsymbol{\textit{y}}$ %  
			\end{minipage}
		}
		\put(1,12){%
			\begin{minipage}{0.4\textwidth}
				\fontsize{9}{14}\selectfont %   
				$\boldsymbol{\textit{y}}$ %  
			\end{minipage}
		}
		\put(15.5,0){%
			\begin{minipage}{0.4\textwidth}
				\fontsize{9}{14}\selectfont %   
				$\boldsymbol{\textit{x}}$ %  
			\end{minipage}
		}
		\put(46,0){%
			\begin{minipage}{0.4\textwidth}
				\fontsize{9}{14}\selectfont %   
				$\boldsymbol{\textit{x}}$ %  
			\end{minipage}
		}
        \put(80,0){%
			\begin{minipage}{0.4\textwidth}
				\fontsize{9}{14}\selectfont %   
				$t/T$ %  
			\end{minipage}
		}
		\put(28.5,14){%
			\begin{minipage}{0.4\textwidth}
				\fontsize{9}{14}\selectfont %   
				$\boldsymbol{\textit{y}}$ %  
			\end{minipage}
		}
		\put(53.5,21.5){%
			\begin{minipage}{0.4\textwidth}
				\fontsize{9}{14}\selectfont %   
				$\omega$ %  
			\end{minipage}
		}
		\put(84,7.5){%
			\begin{minipage}{0.4\textwidth}
				\fontsize{9}{14}\selectfont %   
				$\theta$ %  
			\end{minipage}
		}
		\put(84,5){%
			\begin{minipage}{0.4\textwidth}
				\fontsize{9}{14}\selectfont %   
				$\Omega$ %  
			\end{minipage}
		}
		\put(60.5,15){%
			\begin{minipage}{0.4\textwidth}
				\fontsize{9}{14}\selectfont %   
				$\theta\,({}^\circ)$ %  
			\end{minipage}
		}
		\put(84,21){%
			\begin{minipage}{0.4\textwidth}
				\fontsize{9}{14}\selectfont %   
				$\Omega \,({}^\circ\cdot \mathrm{s}^{-1})$ %  
			\end{minipage}
		}
	\end{overpic}
	\caption{(\textit{a-c}) Swimmer's trajectories for tracking fixed targets (front left, straight and front right). Red dots represent the start and target points. 
     (\textit{d}) The process of making a right turn by the swimmer with action regulation, along with the contour of the vorticity magnitude, at $t/T=0,0.2,0.4,0.7,0.8,1$.  
     The action sequence is $A_1$, $A_3$, $A_0$ and $A_0$. 
     (\textit{e}) Swimmer's trajectory for tracking  the front right target, along with the contour of the vorticity magnitude. 
     (\textit{f}) Temporal evolution of $\theta$ and $\Omega$ for tracking the front right target.}\label{fig:fix_target_alter}
\end{figure}
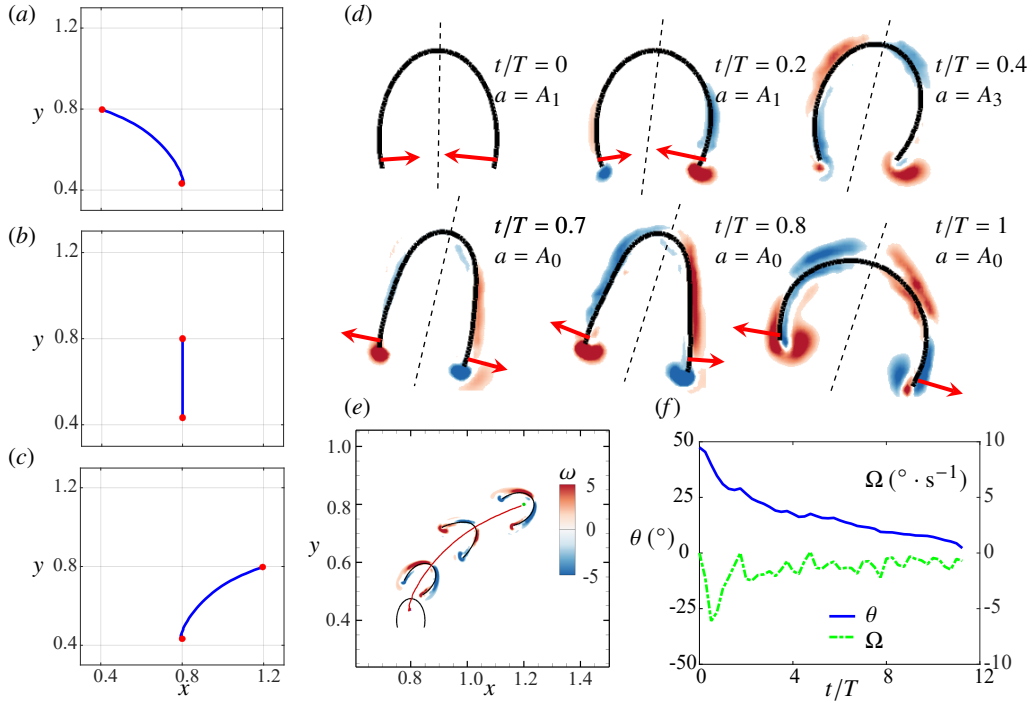

We first examine the DQN performance in tracking a fixed target.
In this case, the swimmer can adjust its movement direction gradually towards the target.
As shown in figures \ref{fig:fix_target_alter}\textit{a-c}, we set three target points.
The swimmer trajectories suggest that the DQN has learned how to make turns using the combination of symmetric and asymmetric actions.

Figures \ref{fig:fix_target_alter}\textit{d} and \textit{e} illustrate how the swimmer moves towards the target point, where the symmetric and asymmetric actions are marked in red and blue, respectively, on the swimmer trajectory.
The first action with asymmetric forces causes the swimmer to turn right by slightly rotating its body, forming a pair of asymmetric wake vortices. 
Under the action regulation (see table~\ref{table:regulation}), the subsequent actions with symmetric forces make the swimmer move straight to the target with a small angular velocity $\Omega$ and a decreasing $\theta$ (see figure \ref{fig:fix_target_alter}\textit{f}), and then make the orientation of its symmetric axis turns slowly to align with its velocity, with the decrease of $\alpha$ (see figure \ref{fig:state_and_geo}\textit{b}).
The decreasing of both $\theta$ and $\alpha$ is ideal for tracking a fixed target, because the swimmer can adjust to proper orientation and then swim forward.
This adjustment is critical, which will be shown in tracking of a moving target.

\subsection{Effect of action regulation}
The detailed process of vorticity formation during the right turn of a swimmer is illustrated in figure \ref{fig:fix_target_alter}\textit{d}, with the sequence of actions $A_1,A_3,A_0$ and $A_0$ in a period from $t=0\,$ to $T$.
First, the force exerted on the right side is larger than that on the left in action $A_1$, generating a stronger vortex at the right trailing edge.
This vortex then induces a leftward rearward velocity on the right half of the swimmer, effectively dragging the entire body towards the right.

The turning process depicted in figure \ref{fig:fix_target_alter}\textit{d} emphasizes the importance of action regulation. By applying forces only in the first and third quarters of each cycle, the swimmer can execute a swift turn within a single period, and the vortices generated in the two quarters have enough time to shed off from the trailing edges within the second and fourth quarters.
Experimental studies showed that the starting vortex at the trailing edge for a impulsively started wing rapidly escapes to the wake within $t/T=0.167$ \citep{Huang2001startvortex}. 
This motion is similar to the flapping by the muscle of the jellyfish-like swimmer. 
Therefore, the time of $T/4$ is enough for these vortices to shed off, with minor influence on lateral motion of the swimmer.

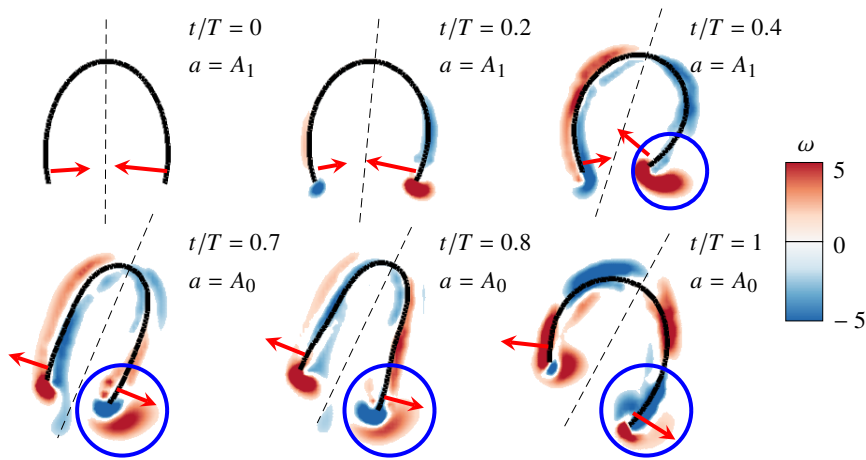
\begin{figure}
	\centering
	\begin{overpic}[scale=0.7]{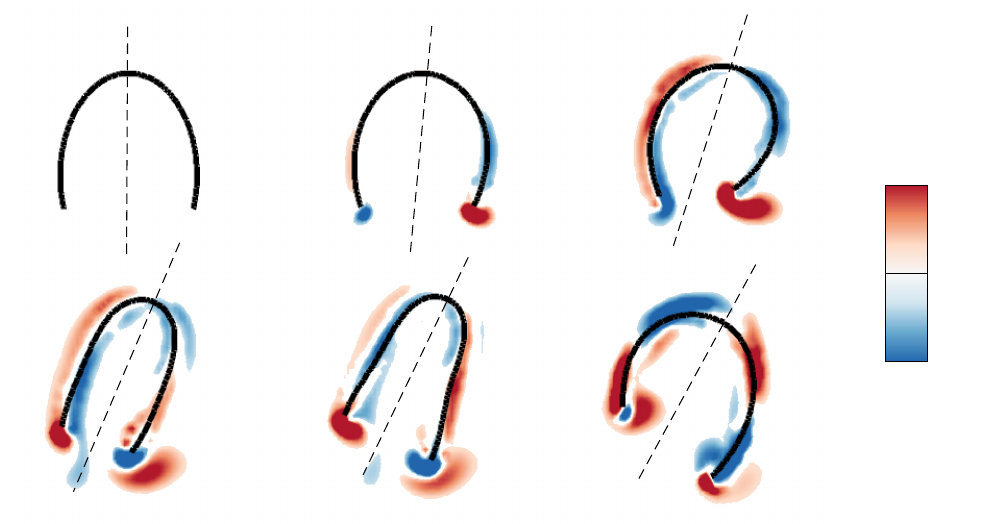}
		\put(78,46){% -17 -3
			\begin{minipage}{0.4\textwidth} %
				\fontsize{9}{14}\selectfont %   
				$t/T=0.4\,\,\, \\ a=A_{1}$ %  
			\end{minipage}
		}
		\put(50,46){%
			\begin{minipage}{0.4\textwidth} %
				\fontsize{9}{14}\selectfont %   
				$t/T=0.2\,\,\, \\ a=A_{1}$ %  
			\end{minipage}
		}
		\put(22,46){%
			\begin{minipage}{0.4\textwidth} %
				\fontsize{9}{14}\selectfont %   
				$t/T=0\,\,\, \\ a=A_{1}$ %  
			\end{minipage}
		}
		\put(22,22){%
			\begin{minipage}{0.4\textwidth} %
				\fontsize{9}{14}\selectfont %   
				$t/T=0.7\,\,\, \\ a=A_{0}$ %  
			\end{minipage}
		}
		\put(50,22){%
			\begin{minipage}{0.4\textwidth} %
				\fontsize{9}{14}\selectfont %   
				$t/T=0.8\,\,\, \\ a=A_{0}$ %  
			\end{minipage}
		}
		\put(78,22){%
			\begin{minipage}{0.4\textwidth} %
				\fontsize{9}{14}\selectfont %   
				$t/T=1\,\,\, \\ a=A_{0}$ %  
			\end{minipage}
		}
        \put(90,35){%
			\begin{minipage}{0.4\textwidth} %
				\fontsize{9}{14}\selectfont %   
				$\omega$ %  
			\end{minipage}
		}
        \put(94,24){%
			\begin{minipage}{0.4\textwidth} %
				\fontsize{9}{14}\selectfont %   
				$5 \\\\ 0 \\\\-5$ %  
			\end{minipage}
		}
		\put(45,27){%
			\begin{tikzpicture}[overlay,scale=0.7]
				\draw[line width=1.5pt, color=red][->](-6.5,1)--(-5.75,1.05);
				\draw[line width=1.5pt, color=red][->](-4.3,1)--(-5.3,1.1);
				\draw[line width=1.5pt, color=red][->](-1.45,1.05)--(-0.9,1.15);
				\draw[line width=1.5pt, color=red][->](0.4,1)--(-0.55,1.2);
				\draw[line width=1.5pt, color=red][->](3.55,1.15)--(4.1,1.25);
				\draw[line width=1.5pt, color=red][->](4.8,1.3)--(4.2,1.8);
				\draw[line width=1.5pt, color=red][->](-6.55,-2.7)--(-7.3,-2.45);
				\draw[line width=1.5pt, color=red][->](-5.25,-3.1)--(-4.5,-3.4);
				\draw[line width=1.5pt, color=red][->](-1.7,-2.45)--(-2.45,-2.15);
				\draw[line width=1.5pt, color=red][->](-0.2,-3.25)--(0.55,-3.45);
				\draw[line width=1.5pt, color=red][->](2.9,-2.3)--(2.0,-2.2);
				\draw[line width=1.5pt, color=red][->](4.5,-3.55)--(5.3,-4.05);
                \draw[color=blue,line width=1.5pt] (5.9,1) arc (0:360:0.7);
				\draw[color=blue,line width=1.5pt] (-4.3,-3.5) arc (0:360:0.85);
				\draw[color=blue,line width=1.5pt] (0.75,-3.5) arc (0:360:0.85);
				\draw[color=blue,line width=1.5pt] (5.6,-3.5) arc (0:360:0.85);
			\end{tikzpicture}
		}
	\end{overpic}
	\caption{The process of making a right turn by the swimmer with action regulation, along with the contour of the vorticity magnitude, at 
        $t/T=0,0.2,0.4,0.7,0.8,1$. The action sequence is $A_1$, $A_1$, $A_0$ and $A_0$, with asymmetric forces in the second quarter. 
        The trailing vortices with opposite signs are generated in different quarters. They cancel out with each other (highlighted in the blue circle), reducing the propulsion efficiency.}\label{fig:vortex_disturb}
\end{figure}

By contrast, frequent exertion of forces can make the swimmer difficult to control.
Without the action regulation, the asymmetric forces can be applied in consecutive quarters as the action sequence $A_1$, $A_1$, $A_0$ and $A_0$ in figure \ref{fig:vortex_disturb}.
As a result, the vortices generated in the consecutive quarters can interference with each other.
In contrast to figure \ref{fig:fix_target_alter}\textit{d}, where the swimmer's geometry becomes nearly symmetric, the vortex (red patch highlighted in the blue circle at $t/T = 0.4$) formed by the right side of the swimmer stays at the trailing edge when the next symmetric action is executed. This interaction between the previously formed vortex and the newly generated vortex leads to an asymmetric shape of the swimmer.
The neighboring vortices  (red and blue patches highlighted in the blue circle at $t/T = 0.7$, 0.8 and 1) with opposite signs cancel out each other, which significantly reduces thrust and makes the motion difficult to control.

\section{Tracking moving target}\label{sec:moving_target}
\subsection{Circular target trajectory tracking}\label{subsec:simple}
We then examine the DQN agent performance in tracking a moving target.
In our training strategy, tracking a moving target is equivalent to tracking a fixed point at each individual time step, so the trained DQN agent is expected to be applicable. 

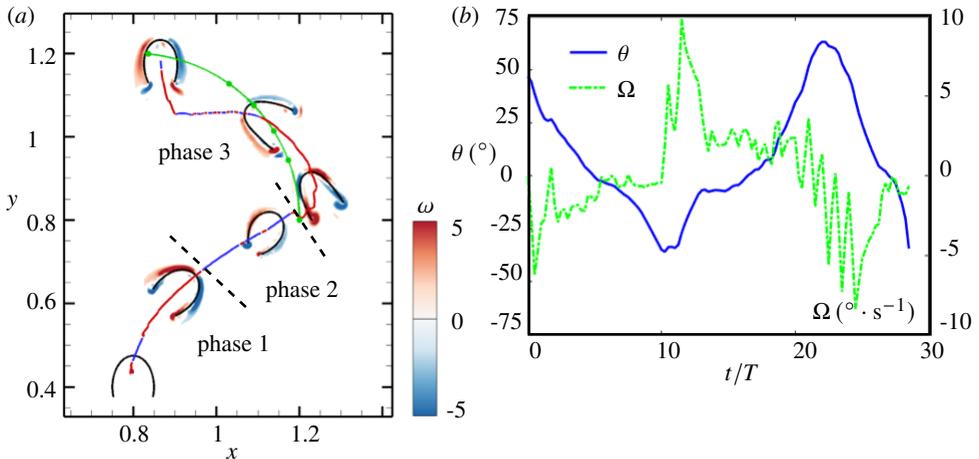
\begin{figure}
	\centering
	\begin{overpic}[scale=0.5]{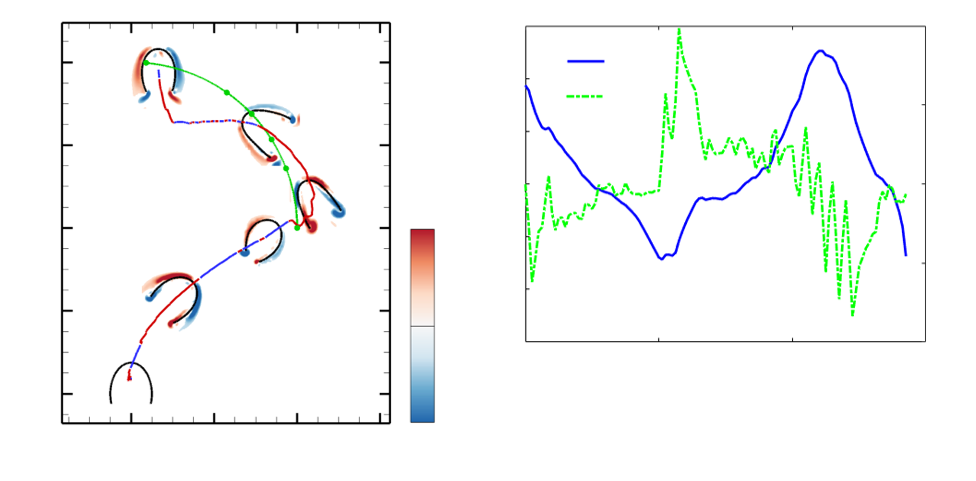}
		\put(0.5,46){%
			\begin{minipage}{0.4\textwidth} %
				\fontsize{9}{14}\selectfont %   
				(\textit{a}) %  
			\end{minipage}
		}
		\put(46,46){%
			\begin{minipage}{0.4\textwidth} %
				\fontsize{9}{14}\selectfont %   
				(\textit{b}) %  
			\end{minipage}
		}
		\put(47,8){%
			\begin{tikzpicture}[overlay,scale=0.7]
				\draw[line width=1pt] (1.28,1.15) rectangle (8.86,7.15);
			\end{tikzpicture}
		}
		\put(50.5,46){%
			\begin{minipage}{0.4\textwidth} %
				\fontsize{9}{14}\selectfont %   
				75 %  
			\end{minipage}
		}
		\put(50.5,41){%
			\begin{minipage}{0.4\textwidth} %
				\fontsize{9}{14}\selectfont %   
				50 %  
			\end{minipage}
		}
		\put(50.5,36){%
			\begin{minipage}{0.4\textwidth} %
				\fontsize{9}{14}\selectfont %   
				25 %  
			\end{minipage}
		}
		\put(51,30){%
			\begin{minipage}{0.4\textwidth} %
				\fontsize{9}{14}\selectfont %   
				0 %  
			\end{minipage}
		}
		\put(50,25){%
			\begin{minipage}{0.4\textwidth} %
				\fontsize{9}{14}\selectfont %   
				-25 %  
			\end{minipage}
		}
		\put(50,20){%
			\begin{minipage}{0.4\textwidth} %
				\fontsize{9}{14}\selectfont %   
				-50 %  
			\end{minipage}
		}
		\put(50,15){%
			\begin{minipage}{0.4\textwidth} %
				\fontsize{9}{14}\selectfont %   
				-75 %  
			\end{minipage}
		}
		\put(54,12){%
			\begin{minipage}{0.4\textwidth} %
				\fontsize{9}{14}\selectfont %   
				0 %  
			\end{minipage}
		}
		\put(67,12){%
			\begin{minipage}{0.4\textwidth} %
				\fontsize{9}{14}\selectfont %   
				10 %  
			\end{minipage}
		}
		\put(81,12){%
			\begin{minipage}{0.4\textwidth} %
				\fontsize{9}{14}\selectfont %   
				20 %  
			\end{minipage}
		}
		\put(94,12){%
			\begin{minipage}{0.4\textwidth} %
				\fontsize{9}{14}\selectfont %   
				30 %  
			\end{minipage}
		}
		\put(74,9.5){%
			\begin{minipage}{0.4\textwidth} %
				\fontsize{9}{14}\selectfont %   
				$t/T$ %  
			\end{minipage}
		}
		\put(96,15){%
			\begin{minipage}{0.4\textwidth} %
				\fontsize{9}{14}\selectfont %   
				-10 %  
			\end{minipage}
		}
		\put(96,22.5){%
			\begin{minipage}{0.4\textwidth} %
				\fontsize{9}{14}\selectfont %   
				-5 %  
			\end{minipage}
		}
		\put(96,30.5){%
			\begin{minipage}{0.4\textwidth} %
				\fontsize{9}{14}\selectfont %   
				0 %  
			\end{minipage}
		}
		\put(96,38.5){%
			\begin{minipage}{0.4\textwidth} %
				\fontsize{9}{14}\selectfont %   
				5 %  
			\end{minipage}
		}
		\put(96,46){%
			\begin{minipage}{0.4\textwidth} %
				\fontsize{9}{14}\selectfont %   
				10 %  
			\end{minipage}
		}
		\put(63,42.3){%
			\begin{minipage}{0.4\textwidth} %
				\fontsize{9}{14}\selectfont %   
				$\theta$ %  
			\end{minipage}
		}
		\put(63,38.6){%
			\begin{minipage}{0.4\textwidth} %
				\fontsize{9}{14}\selectfont %   
				$\Omega$ %  
			\end{minipage}
		}
		\put(42.5,26.5){%
			\begin{minipage}{0.4\textwidth} %
				\fontsize{9}{14}\selectfont %   
				$\omega$ %  
			\end{minipage}
		}
		\put(46,24.5){%
			\begin{minipage}{0.4\textwidth} %
				\fontsize{9}{14}\selectfont %   
				5 %  
			\end{minipage}
		}
		\put(46,15){%
			\begin{minipage}{0.4\textwidth} %
				\fontsize{9}{14}\selectfont %   
				0 %  
			\end{minipage}
		}
		\put(45.5,6){%
			\begin{minipage}{0.4\textwidth} %
				\fontsize{9}{14}\selectfont %   
				-5 %  
			\end{minipage}
		}
		\put(46,32){%
			\begin{minipage}{0.4\textwidth} %
				\fontsize{9}{14}\selectfont %   
				$\theta \, ({}^\circ)$ %  
			\end{minipage}
		}
		\put(83,16){%
			\begin{minipage}{0.4\textwidth} %
				\fontsize{9}{14}\selectfont %   
				$\Omega\,({}^\circ\cdot \mathrm{s}^{-1})$ %  
			\end{minipage}
		}
		\put(47,8){%
			\begin{tikzpicture}[overlay,scale=0.7]
				\draw[line width=1pt][dashed](-5.5,3)--(-4,1.6);
				\draw[line width=1pt][dashed](-3.5,4)--(-2.5,2.5);
			\end{tikzpicture}
		}
		\put(20,12.5){%
			\begin{minipage}{0.4\textwidth} %
				\fontsize{9}{14}\selectfont %   
				phase 1 %  
			\end{minipage}
		}
		\put(27,18){%
			\begin{minipage}{0.4\textwidth} %
				\fontsize{9}{14}\selectfont %   
				phase 2 %  
			\end{minipage}
		}
		\put(16,32){%
			\begin{minipage}{0.4\textwidth} %
				\fontsize{9}{14}\selectfont %   
				phase 3 %  
			\end{minipage}
		}
		\put(12,3.5){%
			\begin{minipage}{0.4\textwidth} %
				\fontsize{9}{14}\selectfont %   
				0.8 %  
			\end{minipage}
		}
		\put(21.5,3.5){%
			\begin{minipage}{0.4\textwidth} %
				\fontsize{9}{14}\selectfont %   
				1 %  
			\end{minipage}
		}
		\put(29.5,3.5){%
			\begin{minipage}{0.4\textwidth} %
				\fontsize{9}{14}\selectfont %   
				1.2 %  
			\end{minipage}
		}
		\put(23,1.5){%
			\begin{minipage}{0.4\textwidth} %
				\fontsize{9}{14}\selectfont %   
				$\boldsymbol{\textit{x}}$ %  
			\end{minipage}
		}
		\put(2.5,8){%
			\begin{minipage}{0.4\textwidth} %
				\fontsize{9}{14}\selectfont %   
				0.4 %  
			\end{minipage}
		}
		\put(2.5,17){%
			\begin{minipage}{0.4\textwidth} %
				\fontsize{9}{14}\selectfont %   
				0.6 %  
			\end{minipage}
		}
		\put(2.5,25){%
			\begin{minipage}{0.4\textwidth} %
				\fontsize{9}{14}\selectfont %   
				0.8 %  
			\end{minipage}
		}
		\put(2.5,33.5){%
			\begin{minipage}{0.4\textwidth} %
				\fontsize{9}{14}\selectfont %   
				1 %  
			\end{minipage}
		}
		\put(2.5,42){%
			\begin{minipage}{0.4\textwidth} %
				\fontsize{9}{14}\selectfont %   
				1.2 %  
			\end{minipage}
		}
		\put(0.5,27){%
			\begin{minipage}{0.4\textwidth} %
				\fontsize{9}{14}\selectfont %   
				$\boldsymbol{\textit{y}}$ %  
			\end{minipage}
		}
	\end{overpic}
	\caption{(\textit{a}) Trajectories of the swimmer's mass centre (red-blue line) and the moving target (green line). The red and blue line segments denote the actions with and without forces, respectively. Six snapshots are presented with the swimmer shape (black curves), vorticity distribution around the swimmer (colour contours), and the target point positions (green dots) at the six time points. 
     (\textit{b}) Temporal evolution of $\theta$ and $\Omega$ for tracking the moving target. The circular target trajectory tracking is also presented in supplementary movie 1.}\label{fig:moving_target}
\end{figure}

The target trajectory (green curve) in figure~\ref{fig:moving_target}\textit{a} is part of a circle centered at $(0.8, 0.8)$ with a radius of $0.4$.
The target point starts from $(1.2, 0.8)$ and moves counterclockwise with an angular velocity of $\pi/120 \, \mathrm{rad}/\mathrm{s}$ or $1.5\, {}^\circ/\mathrm{s}$, which is a speed of $0.0105$ in figure~\ref{fig:moving_target}\textit{a}.

The swimmer effectively tracks the moving target, indicating that the trained neural network is capable of adapting to dynamic scenarios.
Figure \ref{fig:moving_target}\textit{a} and supplementary movie 1 illustrate how the swimmer track a moving target in detail, where the red trajectory denotes the action with forces and the blue one denotes the action without force.
Initially, the target is on the front right of the swimmer.
In phase 1, the swimmer turns right and drifts for a certain period (marked by the blue trajectory) to align its symmetric axis direction with the velocity direction. It then swims straight toward the target (marked by the red trajectory). 
In phase 2, as the target moves to left, the swimmer chooses to drift again (marked by the blue trajectory), allowing it to slow down and adjust its symmetric axis direction towards the moving target.
In phase 3, the swimmer at low speed already turns its symmetric axis toward the target and starts to move forward, and then makes further adjustments to reduce $\theta$.

These observations suggest that the swimmer has acquired the skill of employing drifting (action $A_3$) as a strategy to adjust its direction, enabling it to efficiently follow the moving target. 
Moreover, the large radius of the circular target trajectory provides the swimmer with enough space and time to adjust its swimming direction.
Figure \ref{fig:state_and_geo}\textit{e} sketches the decision process at $t=0$.
The DQN takes the state vector as the input and gives the Q-value for each action, and the action with the largest Q-value is chosen.

Figure \ref{fig:moving_target}\textit{b} plots the temporal evolution of $\Omega$ (angular velocity) and $\theta$ (angle).
In general, $\theta$ and $\Omega$ exhibit opposite signs for the majority of the time, indicating that the swimmer is actively adjusting its direction to swim towards the target.
The delay in the change of $\Omega$ relative to $\theta$ arises from the fact that once an action is performed, it does not instantaneously influence the swimmer's motion.
The swimmer learns to drift, waiting for the previous action to take effect, and then adjust its direction.
Initially, the swimmer employs asymmetric forces to generate an angular velocity, swiftly adjusting its direction towards the target in a short time. Then, the swimmer swims straight forward with minor changes in $\theta$.
When $\theta$ becomes large, the swimmer finds the deviation and responds by producing a counter $\Omega$ to reduce $\theta$.

\subsection{Figure-eight target trajectory tracking}\label{sec:figure_eight}
The trajectory of the target in \S\,\ref{subsec:simple} is a circle with a relatively large radius, moving at a low speed. This provides the swimmer with sufficient time and space to adjust its velocity direction to align with its symmetric axis.

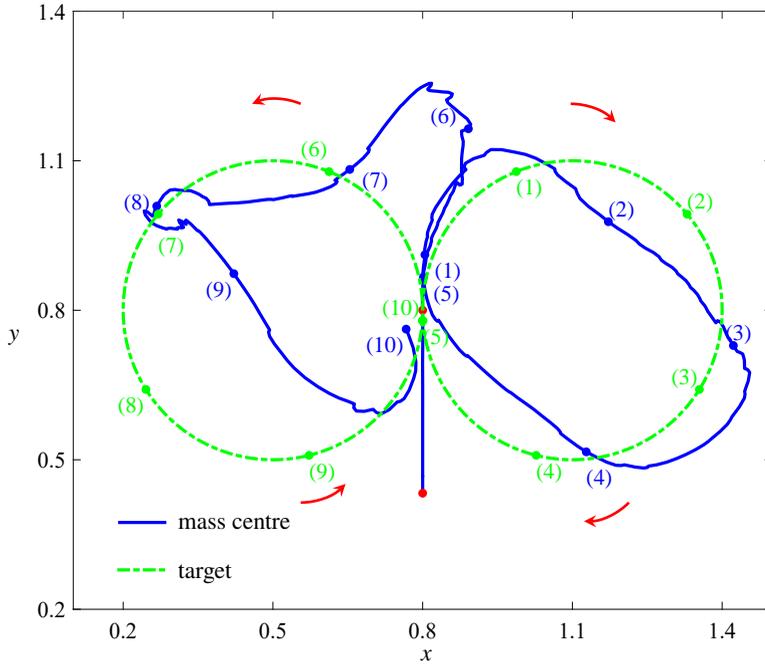
\begin{figure}
	\centering
	\begin{overpic}[scale=0.6]{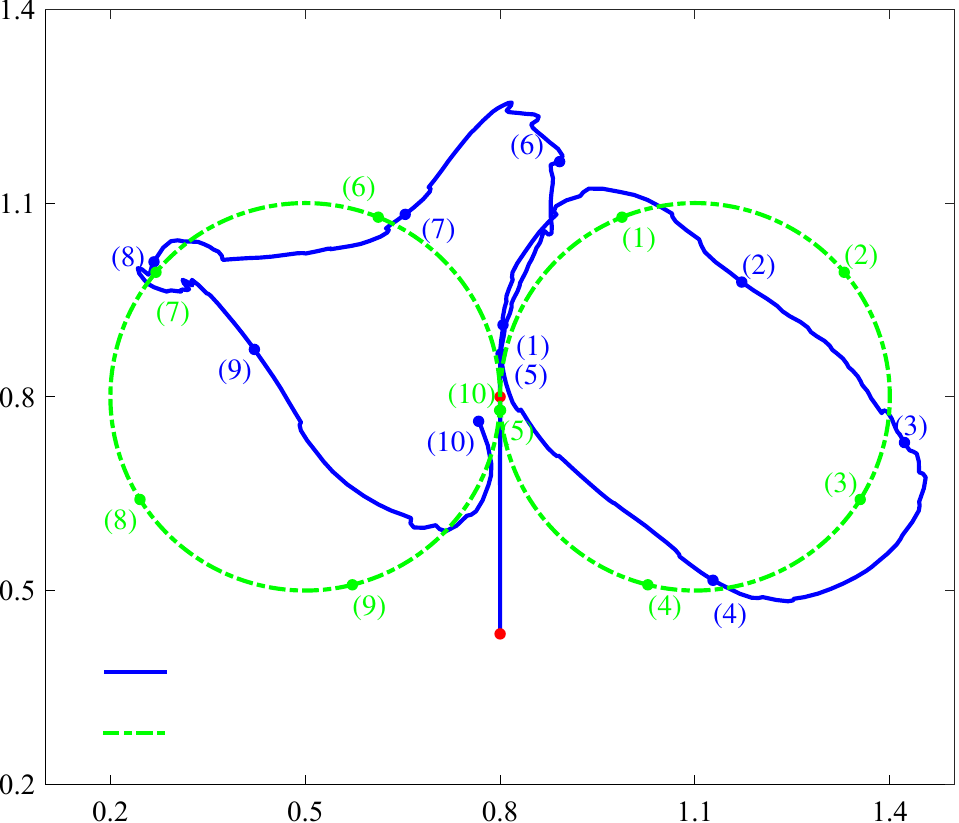} %doubleCircle6.pdf
		\put(19,15){%(21,16)
			\begin{minipage}{0.4\textwidth} %
				\fontsize{9}{14}\selectfont %   
				mass centre %  
			\end{minipage}
		}
		\put(19,8){%(21,9)
		\begin{minipage}{0.4\textwidth} %
			\fontsize{9}{14}\selectfont %   
			target %  
		\end{minipage}
		}
		\put(52,-3){%
			\begin{minipage}{0.4\textwidth} %
				\fontsize{9}{14}\selectfont %   
				$\boldsymbol{\textit{x}}$ %  
			\end{minipage}
		}
		\put(-4,40){%
		\begin{minipage}{0.4\textwidth} %
			\fontsize{9}{14}\selectfont %   
			$\boldsymbol{\textit{y}}$ %  
		\end{minipage}
		}
		\put(50,58){%
		\begin{tikzpicture}[overlay,scale=0.7]
            \draw (3.1,2)[line width=1pt, color=red][->] arc (90:45:1.2);
			\draw (4.2,-5.5)[line width=1pt, color=red][->] arc (-45:-90:1.2);
			\draw (-2,2)[line width=1pt, color=red][->] arc (67.5:112.5:1.2);
			\draw (-2,-5.5)[line width=1pt, color=red][->] arc (-90:-45:1.2);
		\end{tikzpicture}
		}
	\end{overpic}
	\caption{ 
		Trajectories of the swimmer's mass centre (blue line) and the moving target (green line). Red arrows represent the direction of the target's movement. Blue and green dots with different numbers represent the positions of the swimmer and target at the same moments, at $t/T=9,18,...,90$. Red dots represent the starting points of the swimmer and target. The figure-eight target trajectory tracking is also presented in supplementary movie 2.}\label{fig:double_circle}
\end{figure}

Figure \ref{fig:double_circle} and supplementary movie 2 present a more challenging tracking task -- the target moves along a figure-eight trajectory, characterized by a smaller radius and higher speed than those in the previous case. 
Specifically, the radius of each circle is 0.3 and the angular velocity of the target that moves along it is $\pi/45 \, \mathrm{rad}/\mathrm{s}$ or $4\, {}^\circ/\mathrm{s}$. Consequently, the target's speed is $0.021$, which is twice the speed of the target in figure \ref{fig:moving_target}\textit{a}.

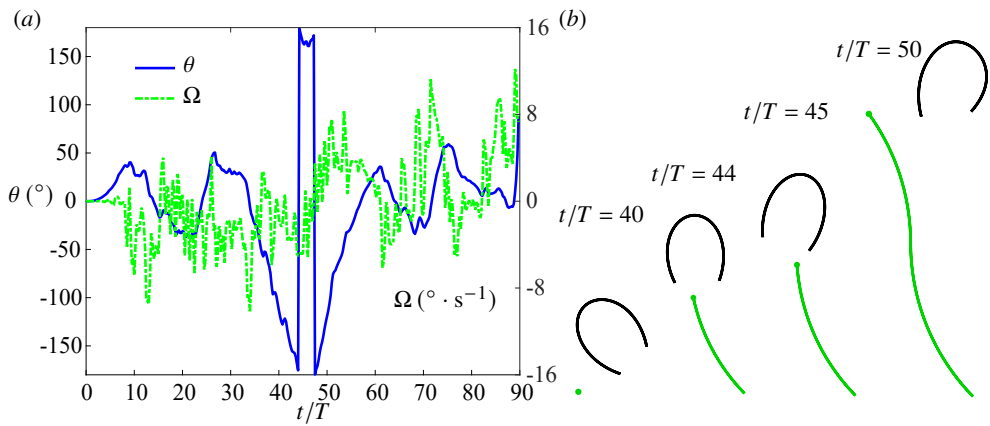
\begin{figure}
	\centering
	\begin{overpic}[scale=0.49]{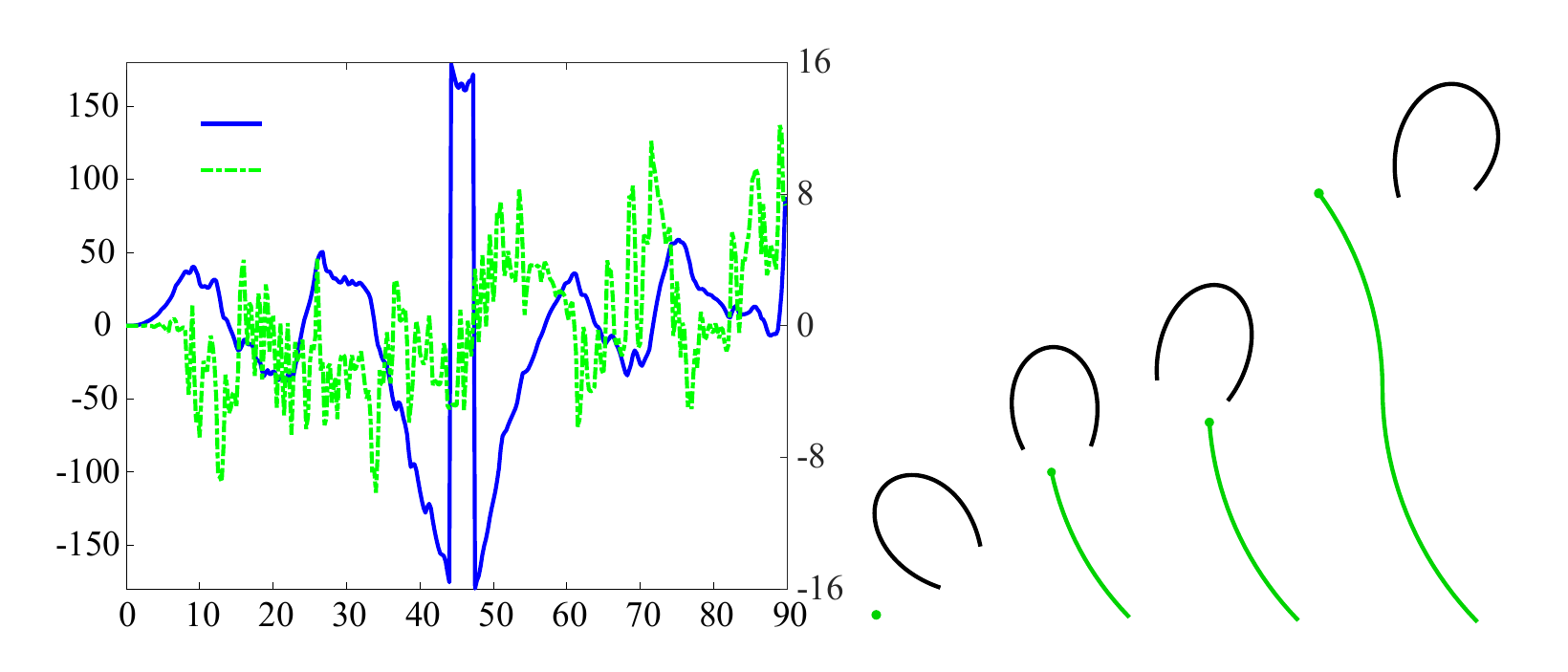}
		\put(17.5,34.7){%
			\begin{minipage}{0.4\textwidth} %
				\fontsize{9}{14}\selectfont %   
				$\theta$ %  
			\end{minipage}
		}
		\put(17.5,31.7){%
			\begin{minipage}{0.4\textwidth} %
				\fontsize{9}{14}\selectfont %   
				$\Omega$ %  
			\end{minipage}
		}
		\put(0.5,22){%
			\begin{minipage}{0.4\textwidth} %
				\fontsize{9}{14}\selectfont %   
				$\theta \, ({}^\circ)$ %  
			\end{minipage}
		}
		\put(38,12){%
			\begin{minipage}{0.4\textwidth} %
				\fontsize{9}{14}\selectfont %   
				$\Omega \, ({}^\circ\cdot \mathrm{s}^{-1})$ %  
			\end{minipage}
		}
		\put(1,39){%
			\begin{minipage}{0.4\textwidth} %
				\fontsize{9}{14}\selectfont %   
				(\textit{a}) %  
			\end{minipage}
		}
		\put(54,39){%
			\begin{minipage}{0.4\textwidth} %
				\fontsize{9}{14}\selectfont %   
				(\textit{b}) %  
			\end{minipage}
		}
		\put(28.5,1){%
			\begin{minipage}{0.4\textwidth} %
				\fontsize{9}{14}\selectfont %   
				$t/T$ %  
			\end{minipage}
		}
		\put(54,20){%
			\begin{minipage}{0.4\textwidth} %
				\fontsize{9}{14}\selectfont %   
				$t/T=40$ %  
			\end{minipage}
		}
		\put(63,24){%
			\begin{minipage}{0.4\textwidth} %
				\fontsize{9}{14}\selectfont %   
				$t/T=44$ %  
			\end{minipage}
		}
		\put(72,30){%
			\begin{minipage}{0.4\textwidth} %
				\fontsize{9}{14}\selectfont %   
				$t/T=45$ %  
			\end{minipage}
		}
		\put(81,36){%
			\begin{minipage}{0.4\textwidth} %
				\fontsize{9}{14}\selectfont %   
				$t/T=50$ %  
			\end{minipage}
		}
	\end{overpic}
	\caption{(\textit{a}) Temporal evolution of $\theta$ and $\Omega$ in the figure-eight target trajectory tracking. Here $\theta$ changes significantly around $t/T=45$ (near station 5 in figure \ref{fig:double_circle}) where the swimmer surpasses the target. 
         (\textit{b}) The swimmer positions relative to the target during it surpasses the target at $t/T=40,44,45,50$. 
         Green dots and lines represent the position and trajectory of the target, respectively. The swimmer first moves close to the target, at which point the target changes its trajectory from clockwise to counterclockwise. This S-shaped maneuver makes it challenging for the DQN agent, which relies solely on the current state, to keep up, resulting in the swimmer temporarily deviating further from the target. 
         }\label{fig:omg_theta_double_circle_overtake}
\end{figure}

Given the delay between the swimmer's motion and action, the swimmer's trajectory does not strictly align with that of the target.
Instead it goes through a process of deviating and adjusting its direction. 
Figure \ref{fig:omg_theta_double_circle_overtake}\textit{a} plots the evolution of $\Omega$ and $\theta$. 
Similar to figure \ref{fig:moving_target}\textit{b}, $\theta$ and $\Omega$ exhibit opposite sign for majority of the time. 
The event for the jumps of $\theta$ around $t/T=45$ is that the swimmer reaches and surpasses the target as shown in figure \ref{fig:omg_theta_double_circle_overtake}\textit{b}. These behaviors indicate that the swimmer's actions are based solely on its current state, without prior knowledge of the target's motion.
Note that the primary objective is to track the target, and the swimmer accomplishes this task for the first time at station 5 in figure \ref{fig:double_circle}. Then, the swimmer restarts to track the moving target.

The trajectory of the swimmer for tracking the first circle (stations 1--5) aligns relatively well with that of the target. 
After reaching the target at station 5, the two trajectories begin to show notable discrepancy. 
Following a period of questionable maneuvers, the swimmer decelerates and manages to reorient itself correctly and catch up at stations 6--8. Difficulties arise at stations 8--9, where the swimmer's velocity turns excessively, leading to further deviation. 
The issues in this complex target trajectory tracking is discussed in detail in Appendix~\ref{sec:complex_traj}. 
Nonetheless, the swimmer ultimately adjusts and catches up at stations 9--10.
Overall, the swimmer still performs satisfactory tracking by completing a rough figure-eight trajectory.

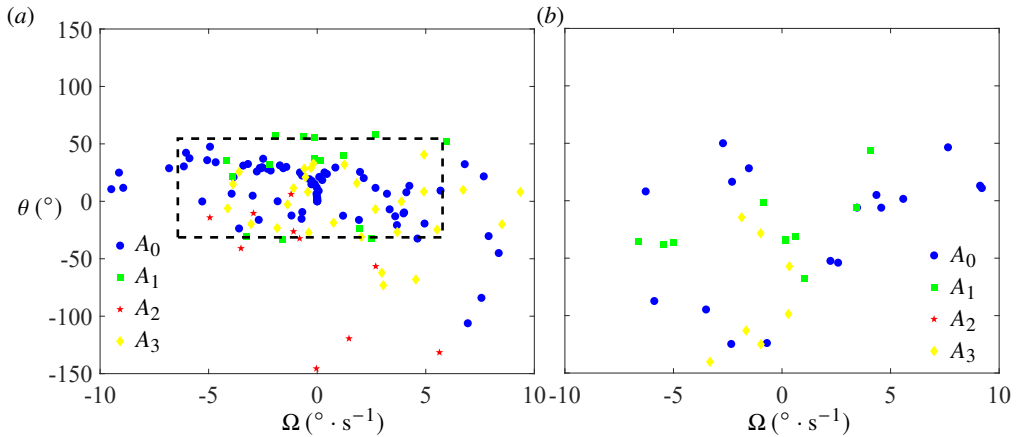
\begin{figure}
	\centering
	\begin{overpic}[scale=0.5]{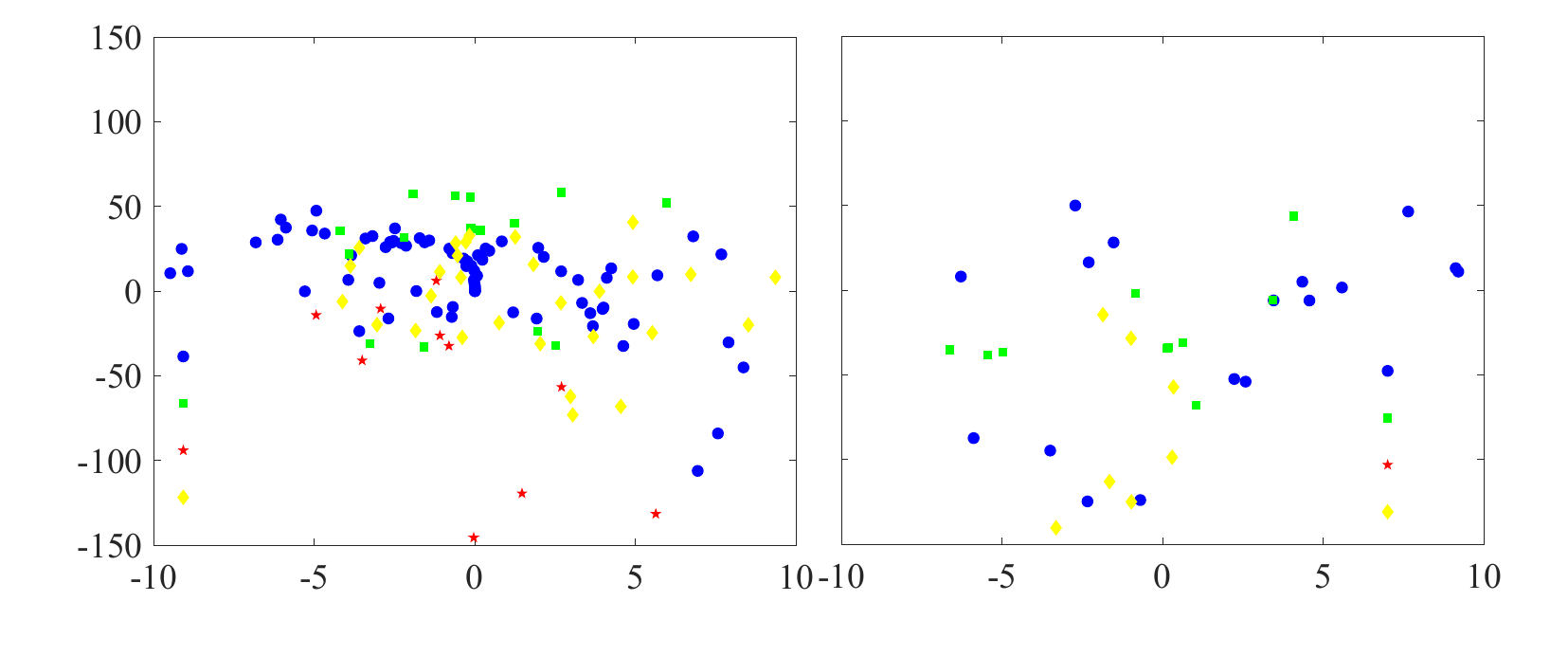}
		\put(1,40){%
			\begin{minipage}{0.4\textwidth} %
				\fontsize{9}{14}\selectfont %   
				(\textit{a}) %  
			\end{minipage}
		}
		\put(51,40){%
			\begin{minipage}{0.4\textwidth} %
				\fontsize{9}{14}\selectfont %   
				(\textit{b}) %  
			\end{minipage}
		}
		\put(2,22){%
			\begin{minipage}{0.4\textwidth} %
				\fontsize{9}{14}\selectfont %   
				$\theta\,({}^\circ)$ %  
			\end{minipage}
		}
		\put(27,2){%
			\begin{minipage}{0.4\textwidth} %
				\fontsize{9}{14}\selectfont %   
				$\Omega\,({}^\circ\cdot\mathrm{s}^{-1})$ %  
			\end{minipage}
		}
		\put(71,2){%
			\begin{minipage}{0.4\textwidth} %
				\fontsize{9}{14}\selectfont %   
				$\Omega\,({}^\circ\cdot\mathrm{s}^{-1})$ %  
			\end{minipage}
		}
		\put(13,18.5){%
			\begin{minipage}{0.4\textwidth} %
				\fontsize{9}{14}\selectfont %   
				$A_0$ %  
			\end{minipage}
		}
		\put(13,15.5){%
			\begin{minipage}{0.4\textwidth} %
				\fontsize{9}{14}\selectfont %   
				$A_1$ %  
			\end{minipage}
		}
		\put(13,12.5){%
			\begin{minipage}{0.4\textwidth} %
				\fontsize{9}{14}\selectfont %   
				$A_2$ %  
			\end{minipage}
		}
		\put(13,9.5){%
			\begin{minipage}{0.4\textwidth} %
				\fontsize{9}{14}\selectfont %   
				$A_3$ %  
			\end{minipage}
		}
		\put(90,17.5){%
			\begin{minipage}{0.4\textwidth} %
				\fontsize{9}{14}\selectfont %   
				$A_0$ %  
			\end{minipage}
		}
		\put(90,14.5){%
			\begin{minipage}{0.4\textwidth} %
				\fontsize{9}{14}\selectfont %   
				$A_1$ %  
			\end{minipage}
		}
		\put(90,11.5){%
			\begin{minipage}{0.4\textwidth} %
				\fontsize{9}{14}\selectfont %   
				$A_2$ %  
			\end{minipage}
		}
		\put(90,8.5){%
			\begin{minipage}{0.4\textwidth} %
				\fontsize{9}{14}\selectfont %   
				$A_3$ %  
			\end{minipage}
		}
		\put(30,27){%
			\begin{tikzpicture}[overlay]
				\draw[line width=1pt][dashed] (-1.8,-1.0) rectangle (1.7,0.3);
			\end{tikzpicture}
		}
	\end{overpic}
	\caption{Scatter plot of $\theta$ and $\Omega$ with different actions output by the DQN agent in the figure-eight target trajectory tracking. 
         The data in the first and third quarters with action regulation are not included. 
         (\textit{a}) Scatter plot for $d\ge d_0$. The data points in the dashed box indicate that the agent frequently selects $A_0$ for small $\theta$. 
         (\textit{b}) Scatter plot for $d< d_0$.}\label{fig:theta_omg_dist}
\end{figure}

We then examine the actions output by the agent. Figure \ref{fig:theta_omg_dist} shows the actions in terms of $\theta$ and $\Omega$. They are divided into two groups, $d<d_{0}$ and $d\ge d_{0}$.
As shown in figure \ref{fig:theta_omg_dist}\textit{a}, as the swimmer maintains a substantial distance from the target, the agent exhibits a pronounced preference for certain actions.
Action $A_0$, with symmetric forcing, predominates for large portion of the time when $\theta$ is small. 
This preference is particularly evident when both $\theta$ and $\Omega$ are small (highlighted by the dashed box in figure \ref{fig:theta_omg_dist}\textit{a}), reflecting the agent prefers straightforward progression with negligible self-rotation under these conditions. 

Action $A_3$ ranks second in the preference, and spans a broader spectrum of $\theta$, corresponding to states characterized by varying velocities. 
This preference implies that the DQN learns a strategy of drifting that enables the swimmer to naturally decelerate or adjust its trajectory by ceasing to apply forces, especially when moving at high speeds in a misaligned direction.
Conversely, $A_1$ and $A_2$ are predominantly selected when $\theta$ has finite positive and negative values, respectively, indicating the swimmer recalibrates its symmetry axis towards the target. 

As shown in figure \ref{fig:theta_omg_dist}\textit{b}, when the swimmer closes in on the target, however, the action selection becomes less predictable. 
Within the vicinity defined by $d<d_0$, the swimmer either approaches immediate proximity to or overshoots the target. 
In this critical range, determining the optimal action becomes increasingly challenging, as any maneuver can precipitate substantial alterations in $\theta$, compounded by the lack of information regarding the target's trajectory. The agent's decision-making process, therefore, becomes more stochastic.

These observations highlight the current challenges of the DQN in handling complex scenarios where precise adjustments and quick responses are required to track a continuously moving target. 
Further fine-tuning of the training process is necessary to overcome these challenges, such as using an extra network to predict the target's trajectory and adding extra action regulations.

\section{Conclusions}\label{sec:conclusions}
We develop a deep reinforcement learning method for tracking a moving target in jellyfish-like swimming.
This control strategy, with a DQN agent, is based on the instantaneous state of a 2D jellyfish-like swimmer.
The swimmer is a flexible object with a muscle model of torsional spring that has no torque introduced when deformed. 
We apply a pair of sinusoidal forces to the muscle part of the swimmer and adjust their magnitudes to control the swimmer's motion.

The data for the swimmer motion in fluid are obtained from the simulation using the immersed boundary method, and they are utilized to train and validate the DQN agent.
To make swimming more natural and reduce the difficulty of training, we introduce an action regulation. 
This regulation mitigates the cancellation of wake vortices generated by the swimmer's flapping motion by suspending the application of forces during specific time periods.

Equipped with the DQN agent and action regulation, the swimmer demonstrates the capability in tracking both fixed and moving targets. 
The action regulation reduces the influence from the flow induced by previous actions, enabling the swimmer to output action to control its movement only based on its current state.	

In the basic test for tracking a fixed target, the swimmer with the DQN agent performs much more efficient than that with the baseline strategy outlined in Appendix \ref{sec:training_details}.
In the challenging test for tracking a moving target, there is the inherent latency between the action for exerting forces and the response for rotating the swimmer's body, due to the hydrodynamic interactions between the shedding vortices and the swimmer's own locomotion. 
The swimmer is still able to dynamically adjust its course based on its instantaneous state. 
This resilience highlights the DQN agent's robust decision-making capabilities, enabling it to counteract external perturbations and maintain a focused pursuit trajectory.

When confronted with more intricate target trajectories, such as the figure-eight curve, the agent's performance may exhibit certain limitations. 
This observation implies the complex interplay of factors that govern natural navigation strategies, particularly those employed by organisms in their native environments. For example, the jellyfish leverages a multifaceted array of sensory inputs, encompassing both internal state cues (e.g.~velocity) and external environmental indicators (e.g.~fluid dynamics) to navigate precisely towards designated locations.

The present study advances the capabilities of a 2D jellyfish-like swimmer beyond forward swimming, explores its control strategy, and broadens the application of reinforcement learning in fluid dynamics. 
Given that the swimmer's motion involves significant fluid-structure interaction, the agent may require additional information to take optimal actions. 
To enhance the tracking performance, the swimmer could leverage more information from its surrounding fluid flow, predict the target's position, and incorporate data from previous states.
Besides, more complex network architectures, such as RNN \citep{rumelhart1986learning} and transformer \citep{transformer2017}, could be employed to capture more features and temporal dependencies from past and current states to achieve higher level of control strategy.

\backsection[Supplementary movie]{\label{SupMat}Supplementary movies 1 and 2 are available.}
\backsection[Acknowledgments]{Numerical simulations were carried out on the TH-2A supercomputer in Guangzhou, China.}
\backsection[Funding]{This work has been supported by the National Natural Science Foundation of China (Grant Nos.~11925201, 12432010 and 11988102), and the Xplore Prize. }
\backsection[Declaration of interests]{The authors report no conflict of interest.}
\backsection[Author contributions]{Y.Y. and Y.C. designed research. Y.C. performed research. All the authors discussed the results and wrote the manuscript. All the authors have given approval for the manuscript.}

\appendix
\section{Elastic force model for the swimmer muscle}\label{sec:elasitc_force}
We present the beam force model for mimicking the muscle of a jellyfish-like swimmer. 
The model swimmer consists of 159 Lagrangian points. Each point is connected to its neighboring ones with linear springs with large stiffness to prevent significant displacement among the Lagrangian points. 
Every set of three neighboring points is connected with a beam that generates a restoring force on those three points when bent, in order to maintain the curvature of the structure.

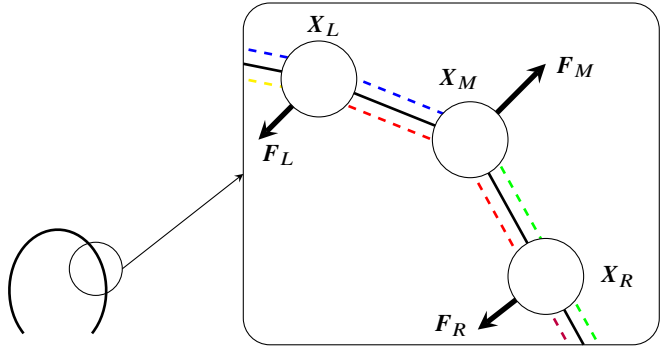
\begin{figure}
	\centering
	\begin{tikzpicture}[scale=1]
		\draw [line width=1pt](-4,-0.55) arc (-45:225:0.64 and 0.8);
		\draw (-3.6,0.3) arc (0:360:0.35);
		\draw [->] (-3.6,0.3)--(-2,1.55);
        \draw [line width=1pt] (-1,2.8)--(1,2);
		\draw [line width=1pt] (1,2)--(2,0.2);
        \draw [line width=1pt] (-1,2.8)--(-2,3);
        \draw [line width=1pt,color=blue][dashed] (-1,3)--(-2,3.2);
        \draw [line width=1pt,color=yellow][dashed] (-1,2.6)--(-2,2.8);
		\draw [line width=1pt] (2,0.2)--(2.5,-0.7);
        \draw [line width=1pt,color=green][dashed] (2.2,0.2)--(2.7,-0.7);
        \draw [line width=1pt,color=purple][dashed] (1.8,0.2)--(2.3,-0.7);
        \draw [line width=1pt,color=red][dashed] (-1,2.6)--(1,1.8);
        \draw [line width=1pt,color=red][dashed] (0.9,1.8)--(1.9,0);
        \draw [line width=1pt,color=blue][dashed] (-1,3)--(1,2.2);
        \draw [line width=1pt,color=green][dashed] (1.1,2.2)--(2.1,0.4);
        \draw[line width=2pt][->](-1,2.8)--(-1.8,2.0);
        \draw[line width=2pt][->](1,2)--(2,3);
        \draw[line width=2pt][->](2,0.2)--(1.1,-0.5);
		\filldraw [fill=white,draw=black] (-1,2.8) circle (0.5);
		\filldraw [fill=white,draw=black] (1,2) circle (0.5);
		\filldraw [fill=white,draw=black] (2,0.2) circle (0.5);
        %\draw[line width=2pt][->](3.5,1)--(3.5,0.3);
		\node[right] at (2.6,0.2){$\boldsymbol{ X}_{R}$};
		\node[above] at (0.85,2.55){$\boldsymbol{ X}_{M}$};
		\node[left] at (-0.6,3.5){$\boldsymbol{ X}_{L}$};
		\node[below] at (2.4,3.2){$\boldsymbol{ F}_{M}$};
		\node[below] at (0.75,-0.2){$\boldsymbol{ F}_{R}$};
		\node[left] at (-1.2,1.8){$\boldsymbol{ F}_{L}$};
        \draw[rounded corners=10pt] (-2,-0.7) rectangle (3.5,3.8);
	\end{tikzpicture}
	\caption{The jellyfish-like model swimmer consists of 159 Lagrangian points. Every pair of adjacent points is connected by a spring, and every three adjacent points are connected by a beam. Black solid lines and dashed lines represent springs and beams, respectively. Different colors correspond to different beams. The discrete fiber beam model of three neighboring points is shown on the right. The three circles represent three neighboring points of a beam model (red dashed line), $\boldsymbol{ X}_{L},\boldsymbol{ X}_{M}$ and $\boldsymbol{ X}_{R}$ the displacement vectors, and $\boldsymbol{ F}_{L},\boldsymbol{ F}_{M}$ and $\boldsymbol{ F}_{R}$ the Lagrangian forces of its left, middle and right point.}
	\label{fig:descrete_structure}
\end{figure}

The structure of the swimmer is sketched in figure \ref{fig:descrete_structure}. The springs and beams together constitute the muscle and geometry of our model swimmer. For the purpose of the present study, the beam needs to resist deformation such as bending, and does not resist translation and rotation so that the whole swimmer can turn without influence from potentially non-zero internal torque.

The elastic force density in \citet{hoover2015jelly} is chosen as
\begin{equation}\label{eq:origin_elastic}
	\boldsymbol {F}_{E}(q,t) = \frac{\partial}{\partial q}\left(k_{s}\left|\frac{\partial}{\partial q}\boldsymbol{ X }(q,t)\right|-1\right)\boldsymbol{\tau}(q,t)- k_{b}\frac{\partial^{4}}{\partial q^{4}}\left(\boldsymbol{ X }(q,t)-\boldsymbol{ X }_{b}(q)\right),
\end{equation}
where $\boldsymbol{ X }(q,t)$ is the Cartesian coordinate of the Lagrangian material point $q$, $\boldsymbol{\tau}(q,t)=(\partial \boldsymbol{ X }/\partial q)/|\partial \boldsymbol{ X }/\partial q|$ is the unit tangent vector, $k_{b}\ge0$ and $k_{s}\ge0$ are respectively the bending and stretching stiffnesses of the muscle, and $\boldsymbol{X}_{b}$ characterizes the shape of the swimmer's body at rest. 
On the right-hand side (RHS) of \eqref{eq:origin_elastic}, the first term represents a linear spring connecting neighboring Lagrangian points, and the second term represents a linear beam applied on three neighboring points (see figure \ref{fig:descrete_structure}).

In the present study, the fiber model for the second term is changed from the linear beam to a torsional spring \citep{IB2d2017} to incorporate rotation, so that the muscle is more flexible than that in \citet{hoover2015jelly}.  
When the entire torsional spring rotates around an arbitrary point, the force acting on it rotates accordingly, while the torque applied to it remains zero.
CFD simulations showed that this enhancement to the muscle model is pivotal for effectively controlling the turning of the jellyfish-like swimmer.

The forces in the torsional spring model are
\begin{eqnarray}\label{eq:torsion}
	\boldsymbol{ F}_{L} = k_{b}C(\boldsymbol{ X}_{M}-\boldsymbol{ X }_{R})\times \boldsymbol{ e}_{z},\\
	\boldsymbol{ F}_{M} = k_{b}C(\boldsymbol{ X}_{R}-\boldsymbol{ X }_{L})\times \boldsymbol{ e}_{z}\label{eq:torsion2},\\
	\boldsymbol{ F}_{R} = k_{b}C(\boldsymbol{ X}_{L}-\boldsymbol{ X }_{M})\times \boldsymbol{ e}_{z},\label{eq:torsion3}
\end{eqnarray}
where $\boldsymbol{ e}_{z} = \boldsymbol{ e}_{x}\times \boldsymbol{ e}_{y}$ is the unit vector perpendicular to the $x$--$y$ plane, $\boldsymbol{ X}_{L},\boldsymbol{ X}_{M}$ and $\boldsymbol{ X}_{R}$ the displacement vector, $\boldsymbol{ F}_{L},\boldsymbol{ F}_{M}$ and $\boldsymbol{ F}_{R}$ the Lagrangian forces of its left, middle and right points, respectively; $C=\boldsymbol{e}_z\cdot((\boldsymbol{X}_R - \boldsymbol{X}_M)\times(\boldsymbol{X}_M - \boldsymbol{X}_L) - (\boldsymbol{X}_{R0} - \boldsymbol{X}_{M0})\times(\boldsymbol{X}_{M0} - \boldsymbol{X}_{L0}))$ and $\boldsymbol{ X}_{L0},\boldsymbol{ X}_{M0}$ and $\boldsymbol{ X}_{R0}$ the corresponding displacement vector at the initial time when the swimmer is at rest. 

The torque referenced to the mass centre $\boldsymbol{ X}_{C} = (\boldsymbol{ X}_{L} + \boldsymbol{ X}_{M}+\boldsymbol{ X}_{R})/3$, or any arbitrary point, is
\begin{equation}\label{eq:torque2}
	\begin{split}
		(\boldsymbol{ X}_{L}-\boldsymbol{ X}_{C})\times\boldsymbol{ F}_{L} + (\boldsymbol{ X}_{M}-\boldsymbol{ X}_{C})\times\boldsymbol{ F}_{M} + (\boldsymbol{ X}_{R}-\boldsymbol{ X}_{C})\times\boldsymbol{ F}_{R}\\
		=\boldsymbol{ X}_{L}\times\boldsymbol{ F}_{L} + \boldsymbol{ X}_{M}\times\boldsymbol{ F}_{M} + \boldsymbol{ X}_{R}\times\boldsymbol{ F}_{R}.
	\end{split}
\end{equation}
Substituting \eqref{eq:torsion}--\eqref{eq:torsion3} into \eqref{eq:torque2} yields the vanishing torque.
This torsional spring model refrains from imparting extraneous torques, notably anti-rotational forces, onto the swimmer's flexible structure. Therefore, it is a suitable model for the present study.

\section{Details on reinforcement learning and training}\label{sec:training_details}

The tracking task is formulated as a sequential decision problem solved using reinforcement learning.
The decision-making is modeled as a Markov decision process where the agent makes decision solely based on the current state \citep{richard2018RL}.

The trajectory of the swimmer can be written as
\begin{equation}
	\Gamma_{t}=(s_{0},a_{0},r_{0},...,s_{t^*-1},a_{t^*-1},r_{t^*-1},s_{t^*},a_{t^*},r_{t^*}),
\end{equation}
where $s_{t^*}$, $a_{t^*}$, and $r_{t^*}$ denote the state, the action taken by the agent, and the reward received at a given time $t=t^*\Delta t$, respectively. In the present study, $\Delta t=T/4$, corresponding to the time interval for force application. 
The agent takes action $a_0$ at $t=0$, and then it receives reward $r_0$ and its state changes from $s_0$ to $s_1$. The same process continues until time $t$.

The action $a_{t^*}$ taken by the agent is determined solely by the current state $s_{t^*}$ and the agent's policy $\pi(\Theta, \cdot)$, which is implemented as a neural network with parameters $\Theta$.
The reward function, $r_{t^*}=r_{t^*}(s_{t^*},a_{t^*})$, depends on the current state and action. 
The goal for our DQN agent is to maximize the expected discounted sum of rewards over time by adjusting $\Theta$ \citep{richard2018RL}. This can be expressed as an optimization problem

\begin{equation}
	\max_{\pi}\mathrm{E}_{s_{0},a_{0},s_{1},a_{1},...}\left(\sum_{t^*=0}^{\infty} \gamma^{t^*}r(s_{t^*},a_{t^*})   \right),
\end{equation}
where $\mathrm{E}_{s_{0},a_{0},s_{1},a_{1},...}$ denotes the expected value taken over the trajectory of states and actions $(s_0, a_0, s_1, a_1, ...)$ that the agent encounters. 
The action $a_{t^*}$ is selected according to policy $\pi$, and $\gamma \in (0, 1]$ is a discount factor that assigns less importance to the reward received in the future. 
We chose this model, considering that the jellyfish has a simple neural structure \citep{jelly2013,jelly2019}, and the current-state information can be sufficient for our model swimmer to navigate.

The training dataset consists of around 40,000 samples called experience tuples. Each sample is described by
\begin{equation}
	(\boldsymbol{s}_{t^*},a_{t^*},r_{t^*},\boldsymbol{s}_{t^*+1},D),
\end{equation}
where $\boldsymbol{s}_{t^*}$, $a_{t^*}$, $r_{t^*}$, $\boldsymbol{s}_{t^*+1}$, and $D$ are the current state, the choice of action, the reward received, the next state, and a flag indicating whether it is the end state of a task, respectively.
Here, $D=1$ denotes the end state when the distance between mass centre and target point is smaller than a certain value $d_{c}$, and $D=0$ otherwise.

Our training dataset is partly computed from a raw dataset, which is obtained by running simulations with randomly chosen actions.
The state vector
\begin{equation}\label{eq:orig_state_vec}
	\begin{split}
		\boldsymbol{s}_{t^*} =(x_{1}^{t^*-1},y_{1}^{t^*-1},x_{2}^{t^*-1},y_{2}^{t^*-1},x_{3}^{t^*-1},y_{3}^{t^*-1},x_{sw}^{t^*-1},y_{sw}^{t^*-1},x_{ta}^{t^*-1},y_{ta}^{t^*-1},n^{t^*-1},\\
		x_{1}^{t^*},y_{1}^{t^*},x_{2}^{t^*},y_{2}^{t^*},x_{3}^{t^*},y_{3}^{t^*},x_{sw}^{t^*},y_{sw}^{t^*},x_{ta}^{t^*},y_{ta}^{t^*},n^{t^*})
	\end{split}
\end{equation}
of samples in the raw dataset is slightly different from that defined in \eqref{eq:state_vec},
where the superscripts $t^*-1$ and $t^*$ denote the data from previous and current time steps, respectively;
$(x_{sw},y_{sw})$ represents the coordinates of the swimmer's mass centre and $(x_{ta},y_{ta})$ the coordinates of the target; 
other variables are the same as in \eqref{eq:state_vec}.

This extended state allows the training dataset to be generated from simulations with multiple different target locations $(x_{ta}, y_{ta})$, and enables the generation of large and diverse training datasets.
The large-scale datasets facilitate the DQN agent learning a robust and adaptable policy for various tracking tasks.

The other variables in \eqref{eq:state_vec} at time $t$ are calculated as follows
\begin{equation}
	\begin{split}
		(u_{1}^{t^*},u_{2}^{t^*})&=\boldsymbol{ \xi}^{t^*} - \boldsymbol{ \xi}^{t^*-1},\\
		\theta^{t^*} &= \mathrm{arccos}\left(\frac{\boldsymbol{ \xi}^{t^*}\cdot\boldsymbol{ \eta}^{t^*}}{|\boldsymbol{ \xi}^{t^*}||\boldsymbol{ \eta}^{t^*}|}\right)\cdot\mathrm{sign}((\boldsymbol{ \xi}^{t^*}\times\boldsymbol{ \eta}^{t^*})\cdot\boldsymbol{e}_{z}),\\
		\Omega^{t^*} &= \theta^{t^*} - \theta^{t^*-1},
	\end{split}
\end{equation}
with
\begin{equation}
	\begin{split}
		\boldsymbol{ \xi}^{t^*} &= (x_{ta}^{t^*},y_{ta}^{t^*})-(x_{sw}^{t^*},y_{sw}^{t^*}),\\		
		\boldsymbol{ \eta}^{t^*}&= (x_{2}^{t^*},y_{2}^{t^*}) - (x_{sw}^{t^*},y_{sw}^{t^*}).
	\end{split}
\end{equation}

\begin{figure}
	\centering
	\begin{overpic}[scale=0.65]{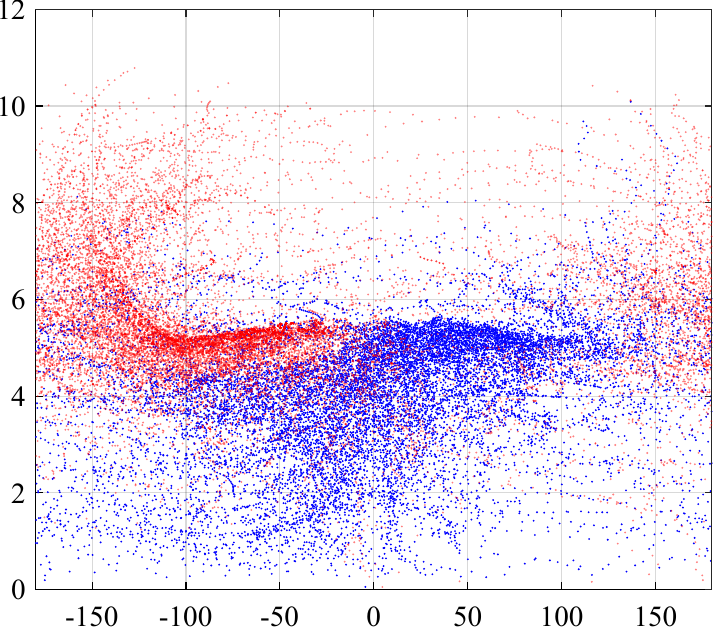}
		\put(-8,41){%
			\begin{minipage}{0.4\textwidth} %
				\fontsize{9}{14}\selectfont %   
				$d/d_0$ %  
			\end{minipage}
		}
		\put(51,-4){%
		\begin{minipage}{0.4\textwidth} %
			\fontsize{9}{14}\selectfont %   
			$\theta\,({}^\circ)$ %  
		\end{minipage}
	}
	\end{overpic}
	\caption{Scatter plot of $d$ and $\theta$. Blue and red dots represent samples from raw dataset and synthetic training dataset, respectively.}\label{fig:d_theta_dist}
\end{figure}

The training dataset consists of two parts. 
One consists of samples directly computed from raw dataset without changing $(x_{sw},y_{sw})$. 
The other one, the synthetic part, is computed with varying $(x_{sw},y_{sw})$ by shifting the target in the raw dataset, and without CFD simulations.  
The $d$--$\theta$ distributions of the samples in the raw and synthetic data in training dataset are shown in figure \ref{fig:d_theta_dist}. 
For reference, the diameter $d_0$ of the jellyfish-like swimmer (distance between the left and right ends of the swimmer at rest) is 0.1 (see table \ref{table:para}). In the raw dataset (blue dots), sample points spread over whole range of $\theta$ and concentrate in $\theta>0^\circ$ with mostly $d/d_0<6$ due to the finite size of the simulation domain. 
To balance this concentration, we vary $(x_{sw},y_{sw})$ so that the synthetic part (red dots) has more samples with $\theta<0^\circ$, which complements the unbalance of $\theta$ in the training dataset. 

The training process is outlined in figure \ref{fig:state_and_geo}. 
The data generated via simulations are utilized in off-line training. The performance of the trained DQN agent is then evaluated. 
Since the convergence of reinforcement learning training is generally not guaranteed, multiple training sessions and evaluations are required.

\begin{figure}
	\centering
	\begin{tikzpicture}[scale=1]
		\draw[line width=1pt] (-2,-0.5) rectangle (0,0.5);
		\node[below] at (-1,0.4) {Input};
		\node[below] at (-1,0.1) {(12D vector)};
		\draw[line width=1pt][->](0,0)--(0.5,0);
		\draw[line width=1pt] (0.5,-1) rectangle (2.1,1);
		\node[below] at (1.3,0.8) {Feature};
		\node[below] at (1.3,0.5) {net};
		\node[below] at (1.3,-0.1) {$(12,64)$};
		\node[below] at (1.3,-0.4) {$(64,64)\times 3$};
		\draw[line width=1pt][->](2.1,0.6)--(3,1.6);
		\draw[line width=1pt][->](2.1,-0.6)--(3,-1.6);
		\draw[line width=1pt] (3,0.5) rectangle (4.6,2.7);
		\node[below] at (3.8,1.1) {$(64,4)$};
		\node[below] at (3.8,1.5) {$(64,64)\times 3$};
		\node[below] at (3.8,2.4) {Advantage};
		\node[below] at (3.8,2) {net};
		\draw[line width=1pt] (3,-2.7) rectangle (4.6,-0.5);
		\node[below] at (3.8,-2.1) {$(64,4)$};
		\node[below] at (3.8,-1.7) {$(64,64)\times 3$};
		\node[below] at (3.8,-0.8) {Value};
		\node[below] at (3.8,-1.2) {net};
		\draw[line width=1pt][->](4.6,1.6)--(5.4,0.6);
		\draw[line width=1pt][->](4.6,-1.6)--(5.4,-0.6);
		\draw[line width=1pt] (5.4,-1) rectangle (6.9,1);
		\node[below] at (6.15,0.5) {Output};
		\node[below] at (6.15,0.2) {Q-values};
		\node[below] at (6.15,-0.1) {(4D vector)};
	\end{tikzpicture}
	\caption{Structure of the dueling network. The number in brackets denotes the number of nuerals for the input and output dimension for a linear layer. Each linear layer follows a ReLU activation layer. The agent's network consists of feature net, advantage net and value net. The output of feature net is the input of the advantage net and value net. The sum of the two nets' output is the final output of the agent, the Q-value for four actions. }\label{fig:network}
\end{figure}
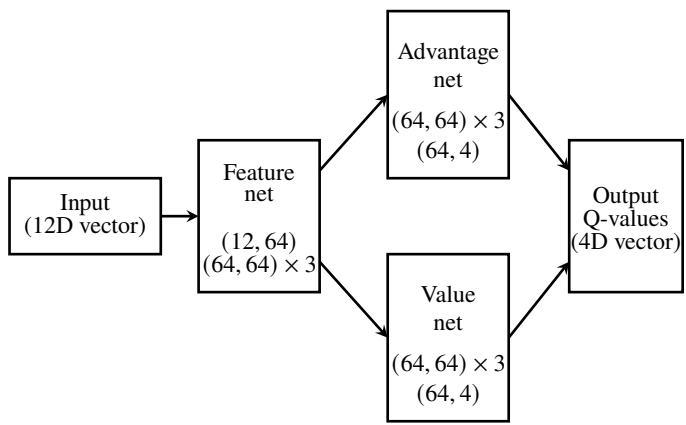

In our implementation, we employed a combination of two variants of the DQN, the duel DQN \citep{wang2016dueling} and the double DQN \citep{van2016deep}, facilitating the fast convergence and training stability. 
The duel DQN is used as the backbone network structure, where the network learns to estimate both the state value function and the advantage function separately. This allows the agent to make informed decisions by comprehending the inherent value of various states and the comparative advantages of different actions.
The double DQN utilizes the policy network and the target network during training. 
The former is responsible for generating Q-values, while the latter is used to compute the target Q-values, helping to mitigate the overestimation of action values and stabilize the training process. The two networks have the same structure as illustrated in figure \ref{fig:network}.

Algorithm~\ref{alg} outlines the procedure for updating the policy and target network parameters throughout the learning process.
Here, the hyper parameters are the discount rate $\gamma=0.99$, and learning rates  $\varepsilon=\varepsilon^{\prime}=10^{-4}$ for deep-Q learning and supervised learning. 
The relaxation factor is $\tau=0.9$.  
Batch sizes for deep-Q learning and supervised learning are both 64.
The loss calculated in line 4 in algorithm~\ref{alg} implements the loss function defined in \eqref{eq:loss}. 
The cross entropy function in line 12 is defined as $H(p,q)=-\sum_{i}p_i\mathrm{log}(q_i)$, where $p_i$ and $q_i$ are target and output distributions, respectively, with $i=0$, 1, 2 and 3. For the supervised learning task here, $q_i=\mathrm{exp}(Q_i)/\sum_i\mathrm{exp}(Q_i)$, where $Q_i$ is the Q-value output of the DQN, and $p$ is defined as a one-hot categorical distribution \citep{bishop2006pattern} 
with $a_t=A_j$ and $p_i=\delta_{ij}$, such as $p=(1,0,0,0)$ for $a_t = A_0$. The delta function is defined as $\delta_{ij}=1$ for $i=j$, and $0$ elsewise.

\renewcommand{\algorithmicrequire}{\textbf{Initialize:}}
\renewcommand{\algorithmicensure}{\textbf{Output:}}
\begin{algorithm}
	\caption{Training of the network}
	\label{EPSA}
	\begin{algorithmic}[1]\label{alg}
		\REQUIRE load dataset obtained from simulation                 
		\REQUIRE policy network $\pi_{p}$ with random parameter $\Theta_{p}$
		\REQUIRE target network $\pi_{t}$ with random parameter $\Theta_{t}$
		\REQUIRE hyper parameters learning rate $\varepsilon$ and $\varepsilon^{\prime}$, 
		discount factor $\gamma$, relaxation factor $\tau$
		\FOR{$i=1$ \TO $N$}
		\STATE $\quad$sample $(\boldsymbol{s}_{t^*},a_{t^*},r_{t^*},\boldsymbol{s}_{t^*+1},D)$ tuple batch from dataset
		\STATE $\quad$calculate every tuple in the batch \\ \quad $Q_{\mathrm{predict}}=r_{t^*}+\gamma(1-D)\mathop{\mathrm{max}}_{A_{j}}  \pi_{p}(\boldsymbol{s}_{t^*+1},A_{j}) $, $Q_{\mathrm{target}}=\pi_{t}(\boldsymbol{s}_{t^*},a_{t^*})$
		\STATE  $\quad$$L=(Q_{\mathrm{predict}}-Q_{\mathrm{target}})^{2}$
		\STATE $\quad$$\Theta_{p}\leftarrow\Theta_{p}-\varepsilon\nabla L$
		\IF{$i\% N_{r}=0$} \STATE $\quad$$\Theta_{t}\leftarrow  \tau\Theta_{p}+(1-\tau)\Theta_{t}$
		\ENDIF
		\IF{$i\%10000=0$}
		\FOR{$k=1$ \TO $1000$}
		\STATE $\quad$load expert tuple batch
		\STATE $\quad$$L^{\prime}=\mathrm{crossEntropy}(\pi_{p}(\boldsymbol{s}_{t^*},a),a_{t^*})$
		\STATE $\quad$$\Theta_{p}\leftarrow\Theta_{p}-\varepsilon^{\prime}\nabla L^{\prime}$
		\ENDFOR
		\ENDIF
		\ENDFOR
	\end{algorithmic}
\end{algorithm}

The choice of the reward function has an impact on the convergence and overall performance of DQN agent. 
The reward function
\begin{equation}\label{eq:reward}
	r(\boldsymbol{s},a)=-\mathrm{min}(\theta^{2},3)+\mathcal A\,\mathrm{clip}(v,-0.1,0.1)-\mathcal B d
\end{equation}
takes into account the swimmer's various behavior.
We set constants $\mathcal A = 20$ and $\mathcal B = 10$ to balance the importance of the three terms in the RHS of \eqref{eq:reward}.  
This reward function encourages that the agent learns to align the swimmer's orientation with the target, maintain an appropriate velocity towards the target, and minimize the distance to the target. 
The first term in the RHS of \eqref{eq:reward} penalizes the squared deviation of $\theta$ from the optimal alignment with the target, up to a maximum value of three. This is an important component of the reward function because the direction of the swimmer's body symmetric axis is crucial. When symmetric forces are applied, the swimmer's velocity quickly adjust to the direction of the symmetric axis.
The second term rewards the swimmer's mass centre velocity towards the target, with the velocity clipped between -0.1 and 0.1 to avoid excessive speed. This term encourages the swimmer to maintain an appropriate velocity towards the target.
The third term minimizes the distance between the swimmer and the target.

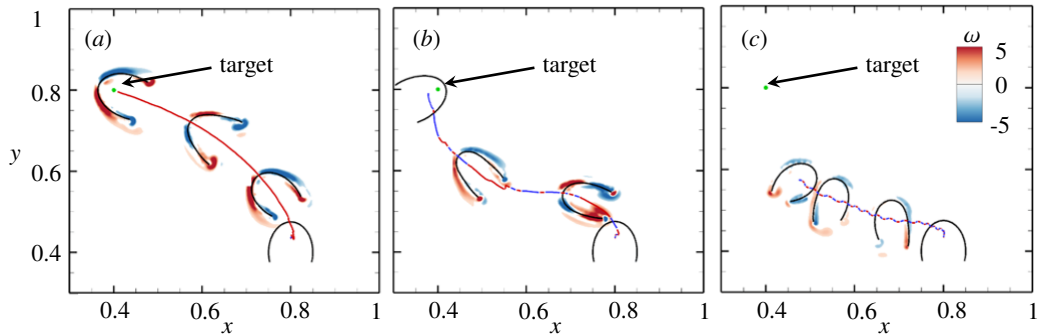
\begin{figure}
	\centering
	\begin{overpic}[scale=0.5]{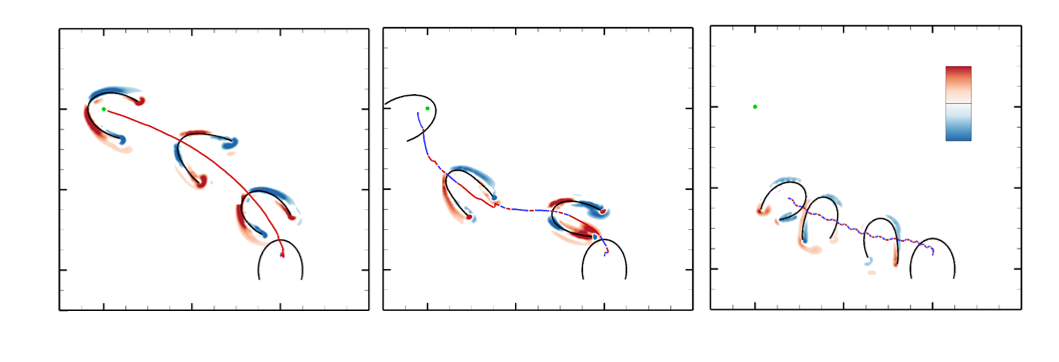}
		\put(7,28){%
			\begin{minipage}{0.4\textwidth} %
				\fontsize{9}{14}\selectfont %   
				(\textit{a}) %  
			\end{minipage}
		}
		\put(38,28){%
			\begin{minipage}{0.4\textwidth} %
				\fontsize{9}{14}\selectfont %   
				(\textit{b}) %  
			\end{minipage}
		}
		\put(69,28){%
			\begin{minipage}{0.4\textwidth} %
				\fontsize{9}{14}\selectfont %   
				(\textit{c}) %  
			\end{minipage}
		}
		\put(14,8){%
			\begin{tikzpicture}[overlay,scale=0.7]
				\draw[line width=1pt][->](1,3.6)--(-0.7,3.3);
				\node[right] at (1,3.6){target};
				\draw[line width=1pt][->](7.1,3.6)--(5.4,3.3);
				\node[right] at (7.1,3.6){target};
				\draw[line width=1pt][->](13.2,3.6)--(11.5,3.3);
				\node[right] at (13.2,3.6){target};
			\end{tikzpicture}
		}
		\put(90,28.3){%
			\begin{minipage}{0.4\textwidth} %
				\fontsize{9}{14}\selectfont %   
				$\omega$ %  
			\end{minipage}
		}
		\put(93,27){%
			\begin{minipage}{0.4\textwidth} %
				\fontsize{9}{14}\selectfont %   
				5 %  
			\end{minipage}
		}
		\put(93,23.8){%
			\begin{minipage}{0.4\textwidth} %
				\fontsize{9}{14}\selectfont %   
				0 %  
			\end{minipage}
		}
		\put(92.5,20){%
			\begin{minipage}{0.4\textwidth} %
				\fontsize{9}{14}\selectfont %   
				-5 %  
			\end{minipage}
		}
		\put(20,1){%
			\begin{minipage}{0.4\textwidth} %
				\fontsize{9}{14}\selectfont %   
				$\boldsymbol{\textit{x}}$ %  
			\end{minipage}
		}
		\put(51,1){%
			\begin{minipage}{0.4\textwidth} %
				\fontsize{9}{14}\selectfont %   
				$\boldsymbol{\textit{x}}$ %  
			\end{minipage}
		}
		\put(82,1){%
			\begin{minipage}{0.4\textwidth} %
				\fontsize{9}{14}\selectfont %   
				$\boldsymbol{\textit{x}}$ %  
			\end{minipage}
		}
		\put(0,17){%
			\begin{minipage}{0.4\textwidth} %
				\fontsize{9}{14}\selectfont %   
				$\boldsymbol{\textit{y}}$ %  
			\end{minipage}
		}
		\put(8.5,2.6){%
			\begin{minipage}{0.4\textwidth} %
				\fontsize{9}{14}\selectfont %   
				0.4
			\end{minipage}
		}
		\put(17,2.6){%
			\begin{minipage}{0.4\textwidth} %
				\fontsize{9}{14}\selectfont %   
				0.6
			\end{minipage}
		}
		\put(25.5,2.6){%
			\begin{minipage}{0.4\textwidth} %
				\fontsize{9}{14}\selectfont %   
				0.8
			\end{minipage}
		}
		\put(34,2.6){%
			\begin{minipage}{0.4\textwidth} %
				\fontsize{9}{14}\selectfont %   
				1
			\end{minipage}
		}
		\put(39.5,2.6){%
			\begin{minipage}{0.4\textwidth} %
				\fontsize{9}{14}\selectfont %   
				0.4
			\end{minipage}
		}
		\put(48,2.6){%
			\begin{minipage}{0.4\textwidth} %
				\fontsize{9}{14}\selectfont %   
				0.6
			\end{minipage}
		}
		\put(56.5,2.6){%
			\begin{minipage}{0.4\textwidth} %
				\fontsize{9}{14}\selectfont %   
				0.8
			\end{minipage}
		}
		\put(65,2.6){%
			\begin{minipage}{0.4\textwidth} %
				\fontsize{9}{14}\selectfont %   
				1
			\end{minipage}
		}
		\put(70.5,2.6){%
			\begin{minipage}{0.4\textwidth} %
				\fontsize{9}{14}\selectfont %   
				0.4
			\end{minipage}
		}
		\put(79,2.6){%
			\begin{minipage}{0.4\textwidth} %
				\fontsize{9}{14}\selectfont %   
				0.6
			\end{minipage}
		}
		\put(87.5,2.6){%
			\begin{minipage}{0.4\textwidth} %
				\fontsize{9}{14}\selectfont %   
				0.8
			\end{minipage}
		}
		\put(96,2.6){%
			\begin{minipage}{0.4\textwidth} %
				\fontsize{9}{14}\selectfont %   
				1
			\end{minipage}
		}
		\put(2,8){%
			\begin{minipage}{0.4\textwidth} %
				\fontsize{9}{14}\selectfont %   
				0.4
			\end{minipage}
		}
		\put(2,15.5){%
			\begin{minipage}{0.4\textwidth} %
				\fontsize{9}{14}\selectfont %   
				0.6
			\end{minipage}
		}
		\put(2,23){%
			\begin{minipage}{0.4\textwidth} %
				\fontsize{9}{14}\selectfont %   
				0.8
			\end{minipage}
		}
		\put(2,30){%
			\begin{minipage}{0.4\textwidth} %
				\fontsize{9}{14}\selectfont %   
				1
			\end{minipage}
		}
	\end{overpic}
	\caption{Comparison of the tracking performance with two DQN agents and the baseline strategy. 
         Trajectories of the swimmer and surrounding vortices are plotted. 
         The red and blue line segments on the swimmer's trajectory denote the actions with and without forces, respectively.     
         Green dot denotes target point. 
         (\textit{a}) Agent 1; (\textit{b}) Agent 2; (\textit{c}) baseline strategy: $a=A_0$ for $|\theta|<\pi/6$, $a=A_1$ for $\theta>\pi/6$, $a=A_2$ for $\theta<-\pi/6$. The total times to complete the tracking task for Agents 1 and 2 are $11.75T$ and $14.25T$, respectively.}\label{fig:compare_base}
\end{figure}

Compared with the standard DQN algorithm, we add an extra supervised learning step into the training process (see lines 9-15 in algorithm \ref{alg}).
This comes from the observation that during training and testing, there are more than one action sequences that can reach the target point. As shown in figures \ref{fig:compare_base}\textit{a} and \textit{b}, Agents 1 and 2 have the DQN trained with different steps to output action.
Although both agents complete the task, Agent 1 has a shorter time.

As shown in figure \ref{fig:compare_base}\textit{c}, we also introduce a synthetic agent as baseline for comparison. 
This baseline strategy has $a=A_0$ for $|\theta|<\pi/6$, $a=A_1$ for $\theta>\pi/6$, and $a=A_2$ for $\theta<-\pi/6$ and also has action regulation.
It is a very straightforward one -- it proceeds straight when the swimmer's direction falls within the tolerance range, and turns left or right when it deviates from that range.
The swimmer with the baseline strategy fails to reach the target. 

\begin{figure}
	\centering
	\begin{overpic}[scale=0.4]{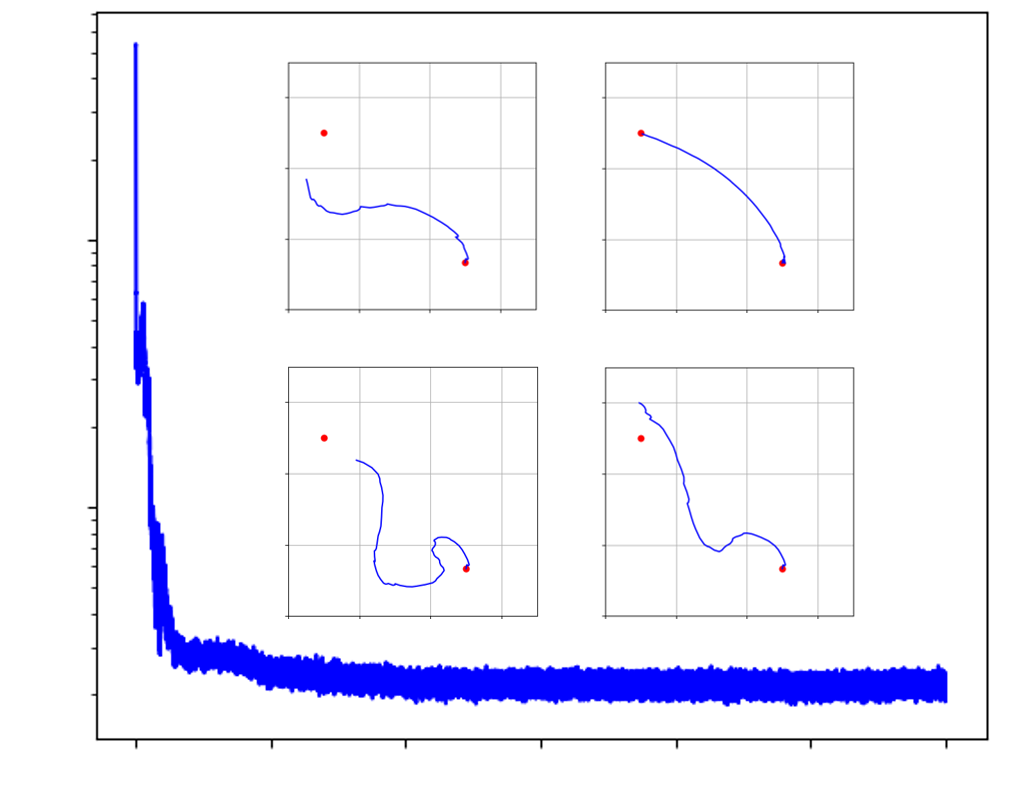}
		\put(22,8){%
			\begin{minipage}{0.4\textwidth} %
				\fontsize{9}{14}\selectfont %   
				A %  
			\end{minipage}
		}
		\put(52,14){%
			\begin{minipage}{0.4\textwidth} %
				\fontsize{9}{14}\selectfont %   
				B %  
			\end{minipage}
		}
		\put(65,14){%
			\begin{minipage}{0.4\textwidth} %
				\fontsize{9}{14}\selectfont %   
				C %  
			\end{minipage}
		}
		\put(91,14){%
			\begin{minipage}{0.4\textwidth} %
				\fontsize{9}{14}\selectfont %   
				D %  
			\end{minipage}
		}
		\put(31,68){%
			\begin{minipage}{0.4\textwidth} %
				\fontsize{9}{14}\selectfont %   
				A %  
			\end{minipage}
		}
		\put(31,38){%
			\begin{minipage}{0.4\textwidth} %
				\fontsize{9}{14}\selectfont %   
				C %  
			\end{minipage}
		}
		\put(62,68){%
			\begin{minipage}{0.4\textwidth} %
				\fontsize{9}{14}\selectfont %   
				B %  
			\end{minipage}
		}
		\put(64,38){%
			\begin{minipage}{0.4\textwidth} %
				\fontsize{9}{14}\selectfont %   
				D %  
			\end{minipage}
		}
		\put(18,8){%
			\begin{tikzpicture}[overlay,scale=0.7]
				\filldraw[color=red] (1.3,0.7) circle (0.18);
				\filldraw[color=red] (5.4,0.45) circle (0.18);
				\filldraw[color=red] (7.5,0.45) circle (0.18);
				\filldraw[color=red] (11.5,0.45) circle (0.18);
			\end{tikzpicture}
		}
		\put(92,2){%
			\begin{minipage}{0.4\textwidth} %
				\fontsize{9}{14}\selectfont %   
				3 %  
			\end{minipage}
		}
		\put(78,2){%
			\begin{minipage}{0.4\textwidth} %
				\fontsize{9}{14}\selectfont 
				2.5 
			\end{minipage}
		}
		\put(66,2){%
			\begin{minipage}{0.4\textwidth} %
				\fontsize{9}{14}\selectfont
				2 
			\end{minipage}
		}
		\put(52,2){%
			\begin{minipage}{0.4\textwidth} %
				\fontsize{9}{14}\selectfont 
				1.5 
			\end{minipage}
		}
		\put(39,2){%
			\begin{minipage}{0.4\textwidth} %
				\fontsize{9}{14}\selectfont 
				1 
			\end{minipage}
		}
		\put(25,2){%
			\begin{minipage}{0.4\textwidth} %
				\fontsize{9}{14}\selectfont 
				0.5 
			\end{minipage}
		}
		\put(13,2){%
			\begin{minipage}{0.4\textwidth} %
				\fontsize{9}{14}\selectfont 
				0 
			\end{minipage}
		}
		\put(2.5,27){%
			\begin{minipage}{0.4\textwidth} %
				\fontsize{9}{14}\selectfont
				$10^{-1}$ 
			\end{minipage}
		}
		\put(86,-1){%
			\begin{minipage}{0.4\textwidth} %
				\fontsize{9}{14}\selectfont %   
				$\times 10^6$ %  
			\end{minipage}
		}
		\put(46,-1){%
			\begin{minipage}{0.4\textwidth} %
				\fontsize{9}{14}\selectfont 
				training steps 
			\end{minipage}
		}
		\put(2.5,54){%
			\begin{minipage}{0.4\textwidth} %
				\fontsize{9}{14}\selectfont 
				$10^{0}$ 
			\end{minipage}
		}
		\put(-2.5,40){%
			\begin{minipage}{0.4\textwidth} %
				\fontsize{9}{14}\selectfont 
				Loss 
			\end{minipage}
		}
	\end{overpic}
	\caption{Convergence of the loss value in one training process without supervised learning, and the corresponding tracking performance (insets A, B, C and D) with $1,1.5,2,2.5,3\times10^{6}$ training steps. As training step increases, the performance of the agent is not guaranteed to become better.}\label{fig:loss}
\end{figure}
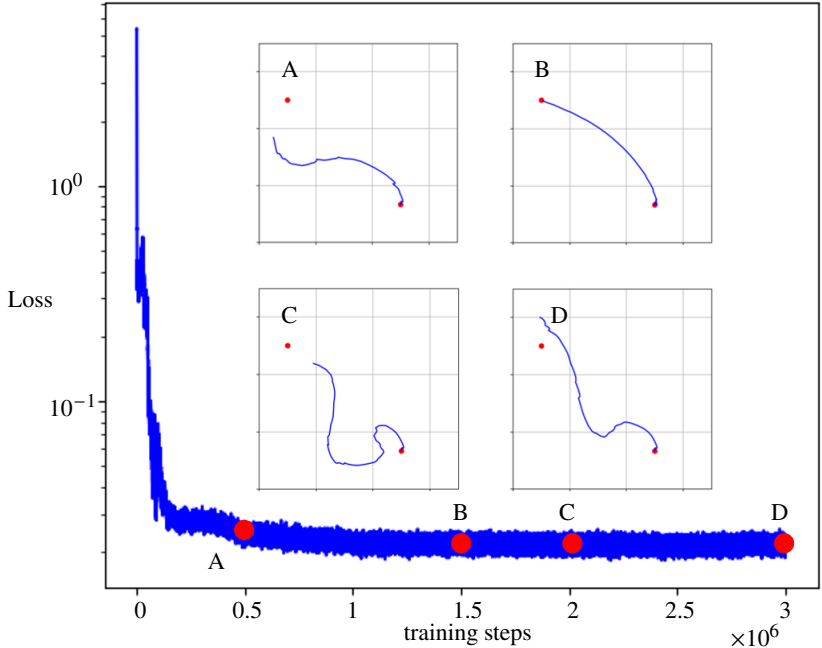

Furthermore, the action sequence provided by one network can be replaced by another in training. 
Among these, we retain the one with the simplest form and shortest duration. 
As depicted in figure \ref{fig:loss}, the network's performance may not consistently improve with the number of training steps, so we employ a supervised learning approach to enforce the network to maintain its choice of actions for certain states.
This training method constrains the network to satisfy the Bellman optimality equation while maintaining known optimal choices for certain states, thereby enhancing the agent's performance. 

\section{Issues in complex target trajectory tracking}\label{sec:complex_traj}

In complex target trajectory tracking in \S\,\ref{sec:figure_eight}, we find that the agent's capability to formulate appropriate actions is compromised when the relative velocity's magnitude $|\boldsymbol{u}_r|$ is substantial and the angle $\alpha$ between $\boldsymbol{u}_r$ and symmetric axis is large (see figure \ref{fig:state_and_geo}\textit{b}).
This deficiency becomes apparent when the target velocity deviates from the swimmer velocity, the swimmer overtakes the target, and the swimmer motion is influenced by the wake flow.

\begin{figure}
	\centering
	\begin{overpic}[scale=0.43]{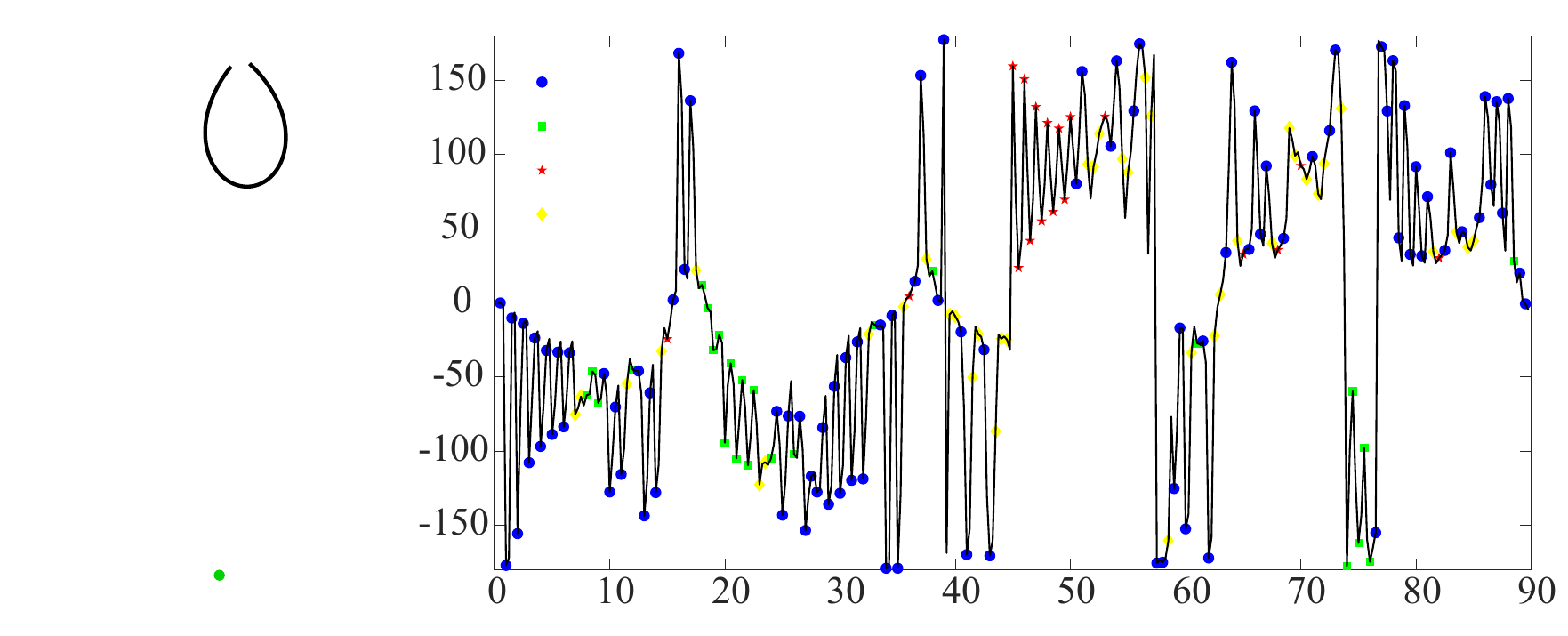}
		\put(6,3){%
			\begin{minipage}{0.4\textwidth} %
				\fontsize{9}{14}\selectfont 
				target 
			\end{minipage}
		}
		\put(64,0){%
			\begin{minipage}{0.4\textwidth} %
				\fontsize{9}{14}\selectfont 
				$t/T$ 
			\end{minipage}
		}
		\put(37.5,22.8){%
			\begin{tikzpicture}[overlay,scale=0.7]
				\draw[line width=1pt,dashed](-4.1,3)--(-3.65,-0.2);
				\draw[line width=1pt,dashed,color=green](-4,2)--(-4.3,-3.45);
				\draw (-4.1,0)[->] arc (-90:-45:0.5);
				\draw[line width=2pt][->](-4,2)--(-5.2,1.9);
				\draw (-4.65,1.85)[->] arc (-170:-60:0.5);

			\end{tikzpicture}
		}
		\put(23,21){%
			\begin{minipage}{0.4\textwidth} %
				\fontsize{9}{14}\selectfont 
				$\alpha \, ({}^\circ)$ 
			\end{minipage}
		}
		\put(22,38){%
			\begin{minipage}{0.4\textwidth} %
				\fontsize{9}{14}\selectfont 
				(\textit{b}) 
			\end{minipage}
		}
		\put(2,38){%
			\begin{minipage}{0.4\textwidth} %
				\fontsize{9}{14}\selectfont 
				(\textit{a}) 
			\end{minipage}
		}
		\put(16,20){%
			\begin{minipage}{0.4\textwidth} %
				\fontsize{9}{14}\selectfont 
				$\theta$
			\end{minipage}
		}
		\put(12,28){%
			\begin{minipage}{0.4\textwidth} %
				\fontsize{9}{14}\selectfont 
				$\alpha$ 
			\end{minipage}
		}
		\put(8,31){%
			\begin{minipage}{0.4\textwidth} 
				\fontsize{9}{14}\selectfont 
				$\boldsymbol{u}_r$ 
			\end{minipage}
		}
		\put(36,34.5){%
			\begin{minipage}{0.4\textwidth} 
				\fontsize{9}{14}\selectfont 
				$A_0$
			\end{minipage}
		}
		\put(36,31.5){
			\begin{minipage}{0.4\textwidth} 
				\fontsize{9}{14}\selectfont 
				$A_1$ 
			\end{minipage}
		}
		\put(36,28.5){
			\begin{minipage}{0.4\textwidth} 
				\fontsize{9}{14}\selectfont 
				$A_2$ 
			\end{minipage}
		}
		\put(36,25.5){
			\begin{minipage}{0.4\textwidth} 
				\fontsize{9}{14}\selectfont  
				$A_3$  
			\end{minipage}
		}
	\end{overpic}
	\caption{
         (\textit{a}) The swimmer position relative to the target at $t/T=72$ (station 8 in figure \ref{fig:double_circle}). 
         (\textit{b}) Evolution of $\alpha$, and the actions chosen at a series of times.}\label{fig:station8}
\end{figure}
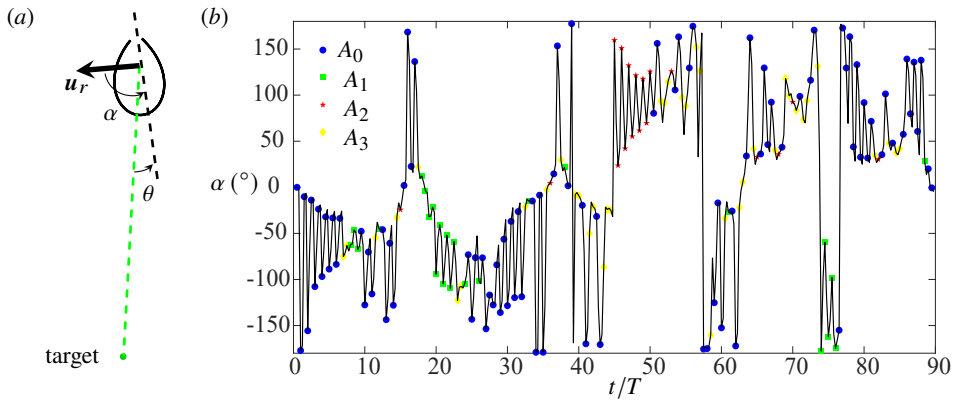

The first scenario emerges when the direction of the target velocity significantly deviates from that of the swimmer. Note that the swimmer's mass centre velocity $\boldsymbol{u}_m$ is different from $\boldsymbol{u}_r$, and they are the same in tracking a fixed target.
Besides, large $\alpha$, even for small $\theta$, also challenges the control of swimmer. This condition is exemplified in figure \ref{fig:station8}\textit{a}, where the swimmer struggles to adjust its course to maintain alignment with the target's path. Although the symmetric axis aligns well toward the target, its $\boldsymbol{u}_r$ deviates, so that the swimmer needs more space and time to reduce the large $\alpha$. 

Figure \ref{fig:station8}\textit{b} shows that action $A_0$ is chosen for majority of the time when $\alpha$ is large, indicating that symmetric force application helps the swimmer to adjust $\boldsymbol{u}_r$ to align with the symmetric axis. 
Note that a large $\alpha$ occurs ultimately because the target can move freely in any direction that deviates from the swimmer velocity. 
When the target velocity deviates from $\boldsymbol{u}_m$, the symmetric axis adjusts its orientation relatively quickly, which leads to large $\alpha$.
These observations imply that optimal tracking performance necessitates not only a minimal $\theta$ but also a small $\alpha$.

The second scenario occurs when the swimmer reaches and surpasses the target, as observed at station 5 in figure \ref{fig:double_circle}. This event triggers a considerable alteration in $\theta$ (see figure \ref{fig:omg_theta_double_circle_overtake}\textit{b}), so that the swimmer is required to execute a sharp U-turn within a confined spatial range to re-establish proximity to the target. 
However, this maneuver is generally unfeasible, because the swimmer is unable to instantly modify its velocity with complex FSI.

\begin{figure}
	\centering
	\begin{overpic}[scale=0.6]{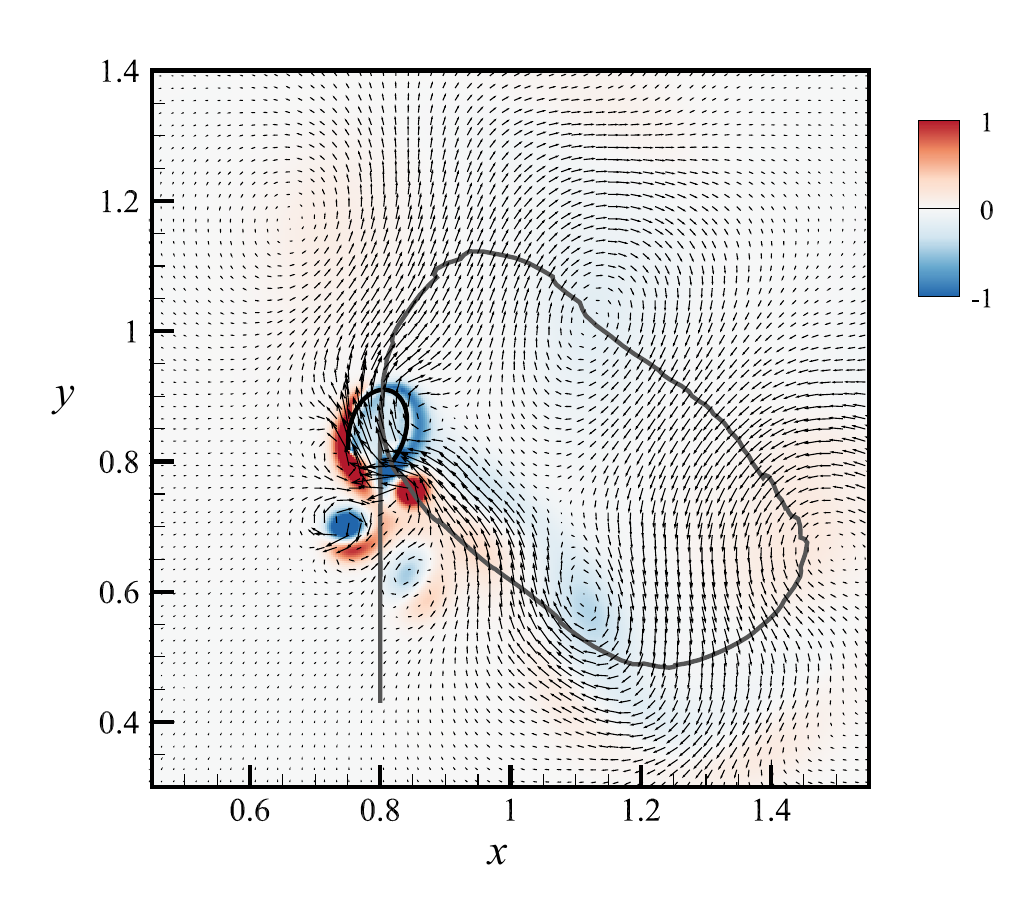}
		\put(91,78){%
			\begin{minipage}{0.4\textwidth} %
				\fontsize{9}{14}\selectfont %   
				$\omega$ %  
			\end{minipage}
		}
	\put(30,37){%
		\begin{tikzpicture}[overlay,scale=0.7]
           \draw[line width=2pt, color=blue][->](-0.6,3.8)--(0.3,2.8);
		\end{tikzpicture}
	}
	\end{overpic}
	\caption{The vorticity contour and velocity field at $t/T=45$ (at station 5 in figure \ref{fig:double_circle}), along with the mass centre trajectory (gray line). The flow induced by previous actions of the swimmer is not negligible. It drags the swimmer to move right upward (marked by the blue arrow), deviating from the figure-eight target trajectory.}\label{fig:prev_ind2}
\end{figure}
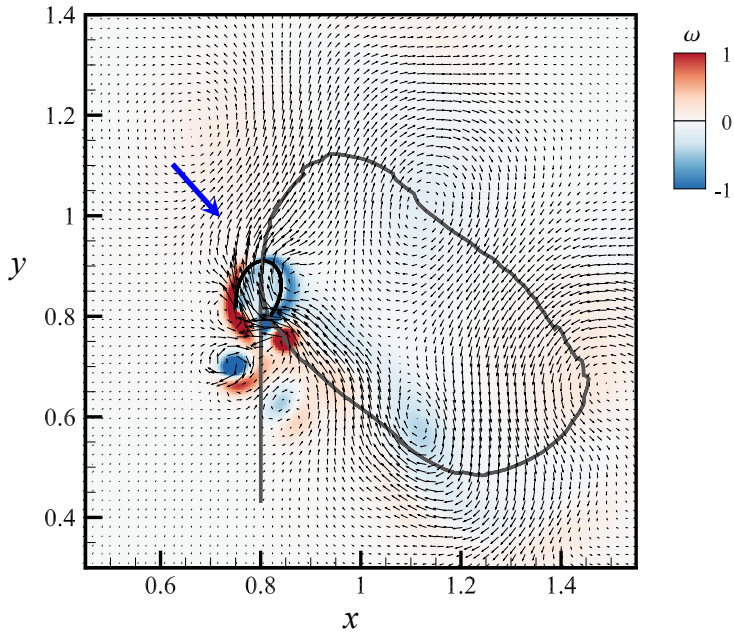

Additionally, the flow induced by earlier movements of the swimmer can disrupt its subsequent motion, particularly when navigating towards the left-side circle. 
Figure \ref{fig:prev_ind2} illustrates the vorticity and velocity field surrounding the swimmer as it approaches its starting position (station 5 in figure \ref{fig:double_circle}). The flow generated by its previous motion exerts a retarding force, interfering the swimmer's directional control.

These issues can hinder the swimmer's ability to promptly and efficiently pursue the target. They underscore a limitation in the current task, wherein the agent operates without anticipating the target's future movements. The unpredictable trajectory of the target poses challenges for the swimmer in adjusting its course accordingly with strong FSI.

\bibliographystyle{jfm}
\bibliography{jelly}

\end{document}